\documentclass[reprint,
superscriptaddress,
 amsmath,amssymb,
 aps,
]{revtex4-1}
\usepackage{amsmath}
\usepackage{graphicx}
\usepackage{dcolumn}
\usepackage{bm}
\usepackage{color}
\usepackage{todonotes}
\usepackage{hyperref}
\usepackage[acronym]{glossaries}

\newacronym{epc}{EPC}{electron-phonon coupling}
\newacronym{xas}{XAS}{X-ray absorption spectroscopy}
\newacronym{rixs}{RIXS}{resonant inelastic x-ray scattering}
\newacronym{arpes}{ARPES}{angle-resolved photoemission spectroscopy}
\newacronym{SL}{SL}{superlattice}
\newacronym{xrd}{XRD}{X-ray diffraction}
\newacronym{lo4}{LO$_4$}{4th longitudinal optical phonon}
\newacronym{sto}{STO}{SrTiO$_3$}
\newacronym{dft}{DFT}{Density Functional Theory}
\newacronym{dos}{DOS}{density of states}

\begin{document}

\title{Decoupling carrier concentration and electron-phonon coupling in oxide heterostructures observed with resonant inelastic x-ray scattering}

\author{D. Meyers}
\email{dmeyers@bnl.gov}
\affiliation{Department of Condensed Matter Physics and Materials Science, Brookhaven National Laboratory, Upton, New York 11973, USA}
\author{Ken Nakatsukasa}
\affiliation{Department of Physics and Astronomy, University of Tennessee, Knoxville, Tennessee 37996, USA}
\author{Sai Mu}
\affiliation{Department of Condensed Matter Physics and Materials Science, Oak Ridge National Laboratory, Oak Ridge, Tennessee 37830, USA}
\author{Lin Hao}
\author{Junyi Yang}
\affiliation{Department of Physics and Astronomy, University of Tennessee, Knoxville, Tennessee 37996, USA}
\author{Yue Cao}
\affiliation{Department of Condensed Matter Physics and Materials Science, Brookhaven National Laboratory, Upton, New York 11973, USA}
\author{G. Fabbris}
\affiliation{Advanced Photon Source, Argonne National Laboratory, Argonne, Illinois 60439, USA}
\author{Hu Miao}
\affiliation{Department of Condensed Matter Physics and Materials Science, Brookhaven National Laboratory, Upton, New York 11973, USA}
\author{J. Pelliciari}
\author{D. McNally}
\author{M. Dantz}
\author{E. Paris}
\affiliation{Research Department Synchrotron Radiation and Nanotechnology, Paul Scherrer Institut, CH-5232 Villigen PSI, Switzerland}
\author{E. Karapetrova}
\author{Yongseong Choi}
\author{D. Haskel}
\affiliation{Advanced Photon Source, Argonne National Laboratory, Argonne, Illinois 60439, USA}
\author{P. Shafer}
\author{E. Arenholz}
\affiliation{Advanced Light Source, Lawrence Berkeley National Laboratory, Berkeley, CA 94720, USA}
\author{Thorsten Schmitt}
\affiliation{Research Department Synchrotron Radiation and Nanotechnology, Paul Scherrer Institut, CH-5232 Villigen PSI, Switzerland}
\author{Tom Berlijn}
\email{tberlijn@gmail.com}
\affiliation{Center for Nanophase Materials Sciences, Oak Ridge National Laboratory, Oak Ridge, TN 37831, USA}
\affiliation{Computational Science and Engineering Division, Oak Ridge National Laboratory, Oak Ridge, Tennessee 37831, USA}
\author{S. Johnston}
\email{sjohn145@utk.edu}
\affiliation{Department of Physics and Astronomy, University of Tennessee, Knoxville, Tennessee 37996, USA}
\affiliation{Joint Institute of Advanced Materials at The University of Tennessee, Knoxville, Tennessee 37996, USA}
\author{Jian Liu}
\email{jianliu@utk.edu}
\affiliation{Department of Physics and Astronomy, University of Tennessee, Knoxville, Tennessee 37996, USA}
\author{M. P. M. Dean}
\email{mdean@bnl.gov}
\affiliation{Department of Condensed Matter Physics and Materials Science, Brookhaven National Laboratory, Upton, New York 11973, USA}

\date{\today}

\begin{abstract}
We report the observation of multiple phonon satellite features in ultra thin superlattices of form $n$SrIrO$_3$/$m$SrTiO$_3$ using resonant inelastic x-ray scattering. As the values of $n$ and $m$ vary the energy loss spectra show a systematic evolution in the relative intensity of the phonon satellites. Using a closed-form solution for the RIXS cross section, we extract the variation in the electron-phonon coupling strength as a function of $n$ and $m$. Combined with the negligible carrier doping into the SrTiO$_3$ layers, these results indicate that tuning of the electron-phonon coupling can be effectively decoupled from doping. This work showcases both a feasible method to extract the electron-phonon coupling in superlattices and unveils a potential route for tuning this coupling which is often associated with superconductivity in SrTiO$_3$-based systems.

\end{abstract}


\maketitle

Despite the discovery of several new classes of superconductors, a comprehensive understanding of superconductivity continues to evade the community, preventing attempts to systamtically control its behavior. The discovery of superconductivity at the interface of two insulating compounds, \gls*{sto} and LaAlO$_3$, is particularly promising for expanding our understanding of superconductivity due to the myriad of control parameters introduced by the heterostructure morphology \cite{Schooley_STO_SC_1964,binnig1980two,ohtomo2004high,reyren2007superconducting}. Furthermore, superconductivity in monolayer FeSe was recently found to be remarkably enhanced by an order of magnitude when interfaced with \gls*{sto} \cite{qing2012interface,liu2012electronic,he2013phase}. These findings point to heterostructuring as a promising route towards the rational engineering of the superconducting ground state.

\begin{figure}[htp]
\includegraphics[width=.5\textwidth]{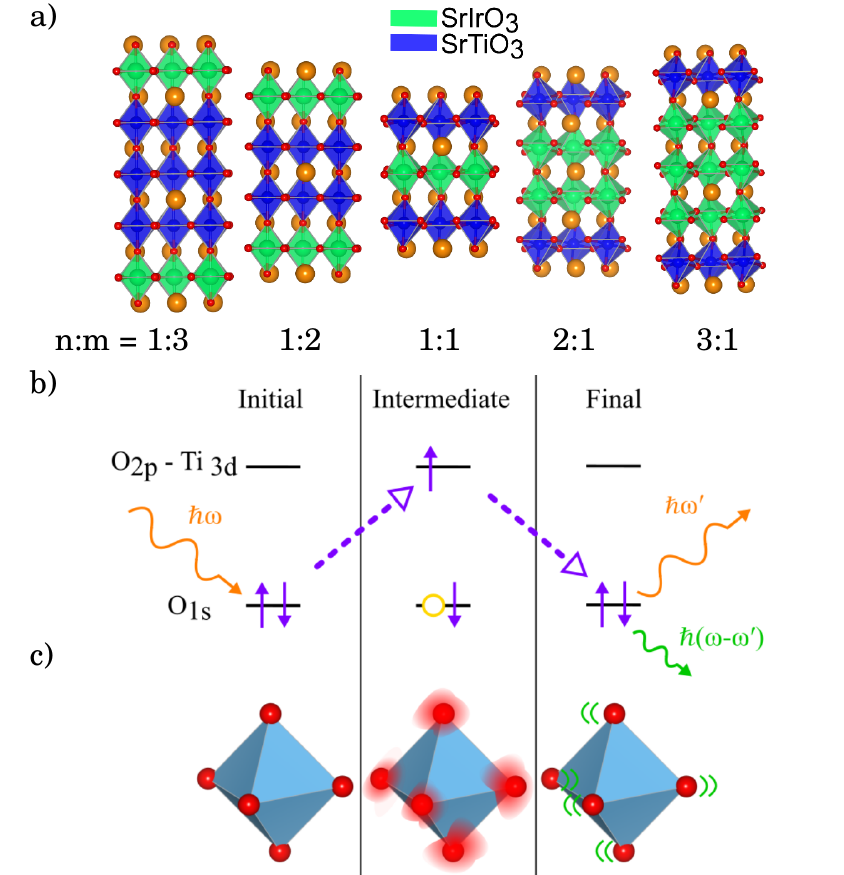} %
\caption{a) Selected \gls*{SL} structures used for this investigation; the octahedral rotations are exaggerated for clarity. b) RIXS process for creating a phonon. c) Example octahedra during the RIXS scattering 
process, with the intermediate state hosting a poorly screened core hole distributed through the lattice that perturbs the O positions. The final state then involves one or more excited \gls*{lo4} phonon modes.
}
\label{Intro}
\end{figure}

While debate remains, the coupling of the conduction electrons to the longitudinal optical (\gls*{lo4}) phonon branch is routinely regarded as an essential ingredient in \gls*{sto}-based superconductors \cite{cowley1964lattice,wang2015tailoring,gervais1993temperature,rosenstein2016superconductivity,baratoff1981mechanism,Swartz2018}. Recent \gls*{arpes} experiments  observed the systematic tuning of the \gls*{epc} through modifications of the carrier density in \gls*{sto} single crystals \cite{wang2015tailoring,moser2013tunable,giustino2017electron}. The superconducting dome of LaAlO$_3$/\gls*{sto} was then conjectured to result from  a delicate balance between the free carrier density and the suppression of polaronic effects with enhanced screening \cite{wang2015tailoring}. Decoupling of the \gls*{epc} from the carrier density by eliminating screening effects and doping associated defects could then fundamentally alter the phase diagram. For example, one might control the \gls*{epc} via interfacial effects while maintaining a fixed carrier density. However, the effect of heterostructuring on the \gls*{epc} is a largely unexplored route, as probing low energy phonon excitations in thin film heterostructures has only recently become possible through advances in Raman scattering and \gls*{rixs}. Traditional methods such as inelastic neutron scattering are unable to attain appreciable signal in the ultrathin film regime ($<50$ nm), while \gls*{arpes} only indirectly couples to phonons and the small electron escape depth means only the first few atomic layers can be effectively probed. In contrast, \gls*{rixs} has penetration depths comparable to the film thickness in the soft x-ray regime and has continuously improved in resolution. Thus, it is currently uniquely situated to measure the low energy excitations in the bulk of ultrathin film superlattices (\gls*{SL}), providing a feasible method to track changes in \gls*{epc} and allowing exploration and scrutiny of another possible dimension of the superconducting phase space \cite{zhou2011localized}.

In this letter, we report the first measurements of \gls*{epc} in ultrathin \gls*{SL} samples of the form $n$SrIrO$_3$/$m$\gls*{sto} utilizing \gls*{rixs} at the O K-edge, extending this method into a new regime \cite{fatale2016hybridization,lee2014charge,johnston2016electron,lee2013role,moser2015electron}. By fitting the phonon excitation profile with a closed form solution for the \gls*{rixs} intensity, we track the systematic change in the \gls*{epc} to the \gls*{lo4} branch as a function of the relative layer thickness \cite{ament2011determining,supplemental}. Such alterations highlight another possible avenue for the engineering of materials unlocked by heteroepitaxy, with important implications for the manipulation of the superconducting state associated with this phonon mode of \gls*{sto} \cite{wang2015tailoring,rosenstein2016superconductivity,he2013phase,lee2013interfacial} and point to the potential of \gls*{rixs} in studying this phenomena \cite{moser2013tunable,lee2013role,moser2015electron,johnston2016electron,devereaux2016directly}.

Samples of form $n$SrIrO$_3$/$m$\gls*{sto} ($n=1,2,3$ with $m=1$ and $m=1,2,3$ with $n=1$) were grown with pulsed laser deposition as detailed elsewhere, Fig. \ref{Intro}(a) \cite{Hao_arxiv}. \gls*{rixs} data was measured with the SAXES spectrometer at the ADRESS beamline of the Swiss Light Source at the Paul Scherrer Institute, with a measured energy resolution of 55 meV at the O K-edge \cite{ghiringhelli2006saxes,schmitt2013high}. All data was taken with scattering vector ${\bf Q} = (0.23, 0, 0.19)$ reciprocal lattice units (r.l.u) and at base temperature $T = 20$~K with an incidence angle of 10$^{\circ}$ corresponding to a $17$ nm sample penetration depth, smaller then the $30-40$ nm sample thickness. \gls*{xas} data for the O K-edge was taken with a total fluorescence yield detector within the sample chamber \footnote{Here r. l. u. is defined within the structural Brillouin Zone with $a$ = $b$ = 3.905, $c$ $\approx$ 3.965.}.  
\begin{figure}[b]
\includegraphics[width=.5\textwidth]{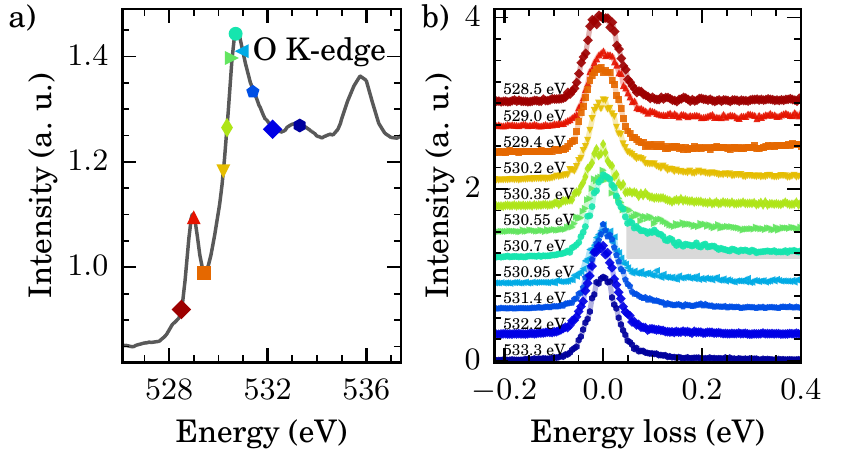} %
\caption{
a) O K-edge XAS taken with vertical polarization for 3/1 SL. Colored markers indicate energy positions where \gls*{rixs} spectra were taken. b) \gls*{rixs} spectra taken at several energies across the absorption edge. The low energy features appearing at the O K-edge white line, 530.7 eV, is highlighted in grey. 
}
\label{Spectra}
\end{figure}

To establish the hybridization with the Ti and Ir \textit{d} states, \gls*{xas} data was taken across the O K-edge, as shown in Fig.~\ref{Spectra}(a). A well-known prepeak feature is clearly visible at $\sim$ 529 eV due to self-doping from the overlap of the O 2$p$ and Ir 5$d$ orbitals  \cite{serrao2013epitaxy}. A white line feature is also observed at $\sim$ 531 eV, and is attributed mostly to the hybridization of the O $2p$ orbitals with the Ti $3d$ orbitals \cite{Groot_OKedge_2017,cao2016orbital}. Higher energy features are also apparent, which signal further hybridization of the O $2p$ orbitals with higher energy Sr, Ir, and Ti orbitals. 

\gls*{rixs} spectra, displayed in Fig.~\ref{Spectra}(b), were taken across the O K-edge at positions indicated by the markers in Fig.~\ref{Spectra}(a). All spectra show an elastic feature of comparable magnitude, centered at zero energy loss. However, clear low energy features are present for incoming x-ray energies tuned to the resonant feature at 530.7 eV, with weight extending out to $\sim 400$~meV energy loss, highlighted in grey. This energy range is typically dominated by lattice, charge, and magnetic excitations. We can rule out magnetic excitations, however, based on the high energy of the features and the resonance being at the white line feature around 530.7 eV that selects the O $2p$ orbitals that are hybridized with the nominally Ti $3d^0$ states, which have no magnetic moments \cite{supplemental,liu2015probing}. Most importantly, as shown below Fig. \ref{Fits}(a), the same features are present in a pure SrTiO$_3$ substrate. Thus, we attribute the low energy features as the signature of multiple phonon excitations. 

\begin{figure}
\includegraphics[width=.5\textwidth]{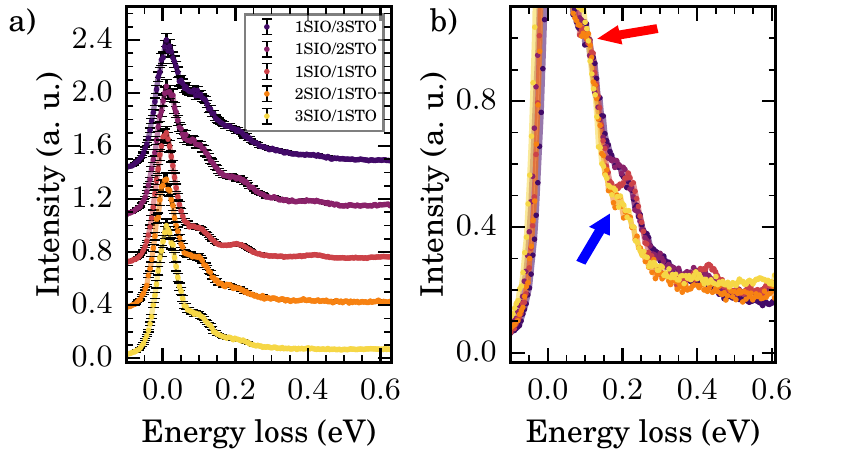} %
\caption{a) RIXS spectra for the five selected \gls*{SL} samples at the O K-edge resonance with phonon features clearly visible, offset vertically for clarity. b) The same spectra normalized to the intensity of first phonon feature at $\sim$ 100 meV, showing the evolution of the relative intensity of the first (red arrow) and second phonon (blue arrow) satellites. 
}
\label{RawRIXS}
\end{figure}

After fixing the incident photon energy to the resonant feature near 530.7 eV, energy loss spectra were taken for the entire series of \gls*{SL} samples [Fig.~\ref{RawRIXS}(a)]. We clearly observe  phonon features in all samples with comparable intensities. As shown previously, in materials with sufficiently strong \gls*{epc}, multiple phonons can be simultaneously excited with \gls*{rixs}  \cite{ament2011resonant,moser2013tunable,lee2013role,johnston2016electron}.   Aligning the spectra and normalizing to the single phonon excitation, Fig. ~\ref{RawRIXS}(b), it becomes clear there is a systematic change in the relative intensity of the multiple phonon features. For the \gls*{SL}s with $n~>~m$, the second feature around 200 meV is weaker. There is, however, a noticeable change in the intensity of the elastic line between samples, likely due to slightly different levels of defects and surface contamination \footnote{To ensure the change in relative intensity of the phonon features is not an artifact of this change, we present the normalized spectra with the fit elastic feature subtracted in the supplemental \cite{supplemental}.}. Despite the differences in the intensity of the elastic feature, the phonon features are sufficiently separated from it (being centered near 100 and 200 meV energy losses) that the influence of the elastic feature is minimal. The observed phonon energy of $\hbar \omega_0 \sim 105$ meV corresponds quite closely to the previously discussed  \gls*{lo4} branch of \gls*{sto} around 100 meV. We assign the observed features as being predominantly from this branch due to the phonon energy, it only appearing at the energy associated with Ti-O bonding while being absent at the Ir-O prepeak \cite{ament2011determining,johnston2016electron}, and the lack of SrIrO$_3$ phonon density of states near this phonon energy (Fig. 5) \cite{gervais1993temperature}. It is clear then that multiple phonon excitations are observed for all samples with a change in the relative intensity of the single and double phonon excitations, which is due to  changes in the EPC \cite{wang2015tailoring,lee2013interfacial,fatale2016hybridization,moser2013tunable}.

\begin{figure*}
\centering
\includegraphics[width=1\textwidth]{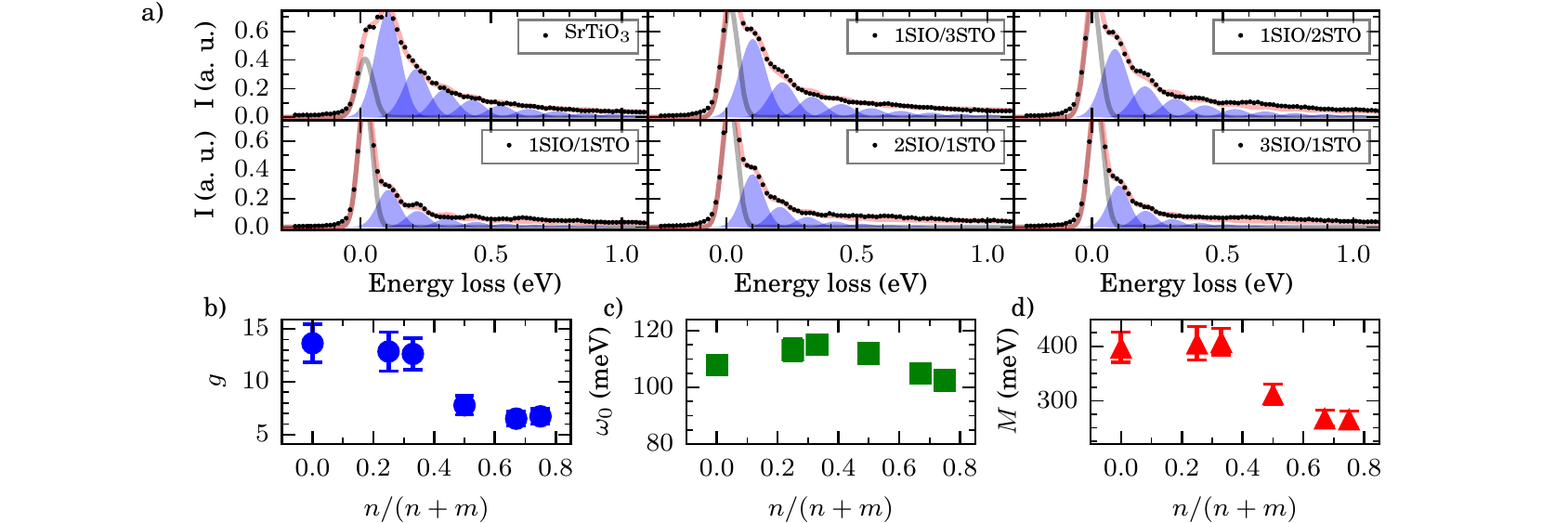} %
\caption{
a) Energy loss spectra at the main O K-edge feature for all samples, including the \gls*{sto} substrate, and the fits to the data. The elastic line is shown in grey, the individual phonon lines in blue, and total fit in red. b) The extracted dimensionless \gls*{epc} strength $g=M^2/\omega_0^2$ obtained from the fits. c) The extracted $\omega_0$ and d) calculated \gls*{epc} constants $M$. 
}
\label{Fits}
\end{figure*}

Analytical solutions to the Kramers-Hesienberg equation for \gls*{rixs} are, in general, very difficult to achieve \cite{ament2011resonant}. However, a closed-form solution has been obtained for the simplified case of a single correlated orbital coupled to a dispersionless Einstein phonon by Ament \textit{et al} \cite{ament2011determining}. In this case, the scattering amplitude of the 
$j^\mathrm{th}$ phonon line $A_{\bf q}(j)$ 
is given by
\begin{equation}\label{Eq:Ament}
\begin{centering}
\begin{aligned}
A_{\bf q}(j) = \sum\limits^{\infty}\limits_{j=0}
\frac{B_{\mathrm{max}(j^{\prime},j),\mathrm{min}(j^{\prime},j)}(g)
B_{k,0}(g)}{\omega_{det}+i\Gamma+(g-k)\omega_0}, 
\end{aligned}
\end{centering}
\end{equation}

where $B_{j,k}(g)=(-1)^j\sqrt{e^{-g}j!k!}\sum\limits^k\limits_{l=0}\frac{(-g)^l\sqrt{g}^{(j-k)}}{(k-l)!l!(j-k+l)!}$ \footnote{Here we use $j$ and $k$ instead of the usual $m$ and $n$ to avoid confusion with the superlattice layering indicies.}. Here, $g = M^2/\omega_0^2$ is a dimensionless measure of the \gls*{epc} energy $M$, $\omega_{det}$ is the energy detuning, $\Gamma$ is the inverse core-hole lifetime, $\omega_0$ is the phonon energy, and $k$ is the index for the phonon eigenstates. Using this equation, the \gls*{rixs} intensity for the phonon modes can be calculated \cite{ament2011determining} and the experimental spectra can then be fit with an elastic feature, the model parameters $\omega_0$, $g$, and $\Gamma$, and a small constant background contribution \cite{fatale2016hybridization,moser2013tunable}. Full details of this fitting routine are provided in the supplemental materials \cite{supplemental}. We note that more complicated cluster calculations \cite{lee2014charge,johnston2016electron,lee2013role} 
produce quantitatively similar phonon excitation profiles as 
the one produced using Eq. \ref{Eq:Ament}, while the former method provides an efficient means to fit the data.

To extract the quantitative changes in the observed relative phonon intensities, we fit the spectra shown in Fig.~\ref{Fits}(a) from all five \gls*{SL} samples and a reference \gls*{sto} substrate using the Ament methodology \cite{ament2011determining}. The fits capture the decay of the phonon intensities up to about 400 meV, after which the intensity becomes too weak to be distinguished from the background. We find that the first four phonon peaks are sufficient to obtain a unique set of fit parameters \cite{supplemental}. Fig.~\ref{Fits}(b-c) show the best fit values for $g$ and $\omega_0$, respectively, plotted as a function of the relative number of SrIrO$_3$ layers to the total number of layers, $\frac{n}{n + m}$. The phonon energy varies within a range only slightly larger then the error bars of approximately $\pm 5$ meV. These two parameters can be combined to obtain the \gls*{epc} energy $M=\sqrt{\omega_0^2g}$, as shown in Fig.~\ref{Fits}(d). For all samples with $n<m$ (i.e. majority Ti-layers), the coupling energy is quite flat with $M\approx400~$meV, as is seen for the bulk \gls*{sto} sample. This value is similar in magnitude to that obtained for other titanates \cite{fatale2016hybridization,Iguchi_TheoryBTO_EPC}. For $n=m$, however, there is a large, abrupt drop of $\sim 20\%$, and a further drop $\sim 10\%$ for $n>m$. Interestingly, the coupling here appears to have stabilized, with no further decrease between $n=2$ and $3$, implying that the microscopic mechanism governing the \gls*{epc} has saturated.

The \gls*{epc} can be modified through changes in carrier density, which affects the degree of electronic screening \cite{wang2015tailoring,moser2013tunable,giustino2017electron}. Measurements of doped \gls*{sto} with \gls*{arpes} revealed the \gls*{epc} could be strongly reduced with changes to the Ti valence of approximately $0.1-0.2~e^-$/Ti \cite{wang2016tailoring}. Such a change could conceivably be induced in our samples through charge transfer at the interface, as seen in other perovskite heterostructures  \cite{gray2016superconductor,cao2016engineered,zhong2017band,Hwang_ReviewSL2012,Chakhalian_ReviewSL2014,mannhart2010oxide}. However, this scenario is difficult to rationalize in light of the rather sudden onset of the change in $g$ observed here and the lack of a change for samples with $n<m$ compared to the bulk \gls*{sto}. The lack of interfacial charge transfer is also expected based on theoretical calculations, which show the lowest lying Ti $3d$ bands are $\sim$0.5-1.0 eV from the top of the valence band \cite{Matsuno2015_SIOSTO,fan2015_LDA,Kim2017_STSIOtheory}. Furthermore, transport and x-ray absorption spectroscopy do not indicate any appreciable doping, and no deviations from Ti$^{4+}$ and Ir$^{4+}$ have been observed previously \cite{spinelli2010electronic,moos1996electronic,Hao_arxiv,Matsuno2015_SIOSTO}. Recent \gls*{rixs} work on TiO$_2$ with carrier doping of 0.01 e$^-$/Ti failed to show a noticeable effect on the measured \gls*{epc}, indicating a much larger doping would likely be needed to induce the observed changes \cite{moser2013tunable}. Thus, enhancement of electronic screening with increased carrier concentration appears an unlikely source for the observed change of the \gls*{epc}.

\begin{figure}[b]
\includegraphics[width=.5\textwidth]{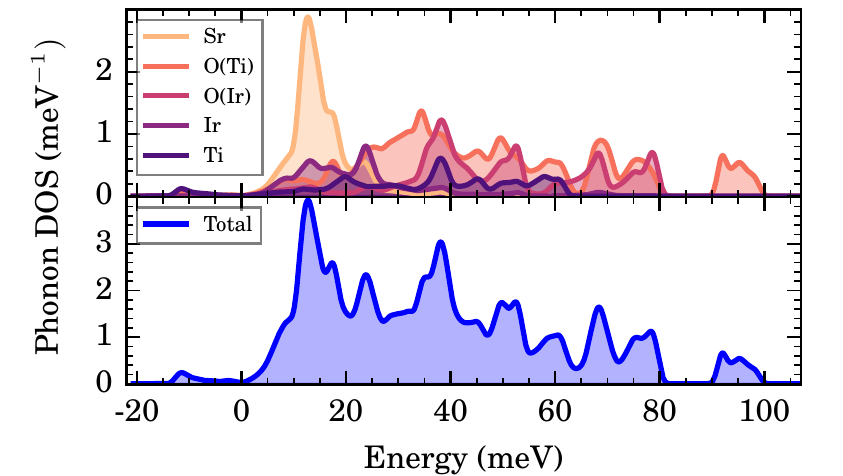} %
\caption{
Partial phonon \gls*{dos} of the $1$SrIrO$_3$/$1$SrTiO$_3$ superlattice obtained from Density Functional Theory (upper panel). Here O(Ti) are defined as the O atoms in the TiO$_2$ and SrO layers, which are shared with the Ir octahedra, and O(Ir) are defined as the O atoms laying in the IrO$_2$ planes. Total \gls*{dos} is shown in the lower panel. Phonon DOS is only present around 100 meV for O atoms associated with the SrTiO$_3$ layers.}
\label{figure4}
\end{figure}

To better understand these observations, we performed \gls*{dft} calculations \cite{supplemental}. Figure~\ref{figure4} displays the partial phonon \gls*{dos} of the $n=m=1$ superlattice, where the structure was taken from Ref.~\cite{khkim_2014}. The highest LO branch $\sim$ 95 meV contains the modes producing the phonon excitations observed in RIXS. The calculation shows that these modes are practically fully residing in the STO layers, despite the heterostructering with SIO, which simplifies the analysis.

Next, to quantitatively evaluate and distinguish possible mechanisms for the observed modulations of the EPC we consider two scenarios. The first scenario is that the EPC is dominated by the Fr\"{o}hlich mechanism, where the LO phonons couple to the electrons via the macroscopic electric fields perturbed by the atomic vibrations~\cite{frohlich_1954}. From our calculations, we find that the SIO layers are less polar than the STO layers. For example, the Born effective charges of the O (Ir) atoms in the SIO layer are a factor 1.4 (1.8) smaller than those of the O (Ti) atoms in the STO layer~\cite{supplemental}. This is probably because STO, unlike SIO, has an empty d-shell, and because the Ir-$5d$ orbitals are more extended and more covalently bonded with their ligand O atoms than the Ti-$3d$ orbitals~\cite{matthias_1949,nicola_2000}. Furthermore, we find the high-frequency dielectric constant $\epsilon_{\infty}$ in the SL increases by a factor of 1.9 compared to bulk STO~\cite{supplemental}. Both a reduced polarity and an increased $\epsilon_{\infty}$ will weaken the Fr\"{o}hlich coupling. This scenario is  consistent with the observed reduction of the EPC in going from bulk STO to the 1/1 SL. However, to rigorously validate this idea, the EPC needs to be computed taking into account the interactions and spin-orbit coupling in the Ir-5$d$ shell and the dynamic screening, including lattice contributions ~\cite{giustino_rmp_2017,Klimin2017}. Such an analysis is beyond the scope of the current manuscript. 

The second scenario we consider is that the octahedral distortions differ from one layer to another, modifying the EPC in the different SLs. Experimentally, we found the $n=1,2,3$ samples do exhibit octahedral rotations about the $a$-$b$ axes that are not observed in the other SLs \cite{supplemental}. However, separate \gls*{dft} calculations performed on bulk-like \gls*{sto} with such behavior find only a weak and opposite trend, i.e. larger $a$-$b$ axis rotations slightly increase the EPC, at odds with our RIXS result \cite{supplemental}. We note that the rotations used in our \gls*{dft} calculations are based on XRD refinements in which precise rotation angles are challenging to exactly resolve in thin film heterostructures, but the effect of small deviations is not expected to provide meaningful changes to the analysis.

Despite this, the failure of this scenario to theoretically induce such changes in the \gls*{epc} point to the Fr\"{o}hlich mechanism as dominating the observed changes. This result provides promise for utilizing different spacing layers with varied polarity and dielectric properties to tune the \gls*{epc} in thin film SLs, opening another dimension to explore the phase diagram in \gls*{sto}-based superconductors. We note the next generation of \gls*{rixs} spectrometers with highly improved energy resolution will make this technique quite practical in the future, allowing \gls*{epc} in single layer films and phonons with lower $\omega_0$ to be efficiently probed and the $Q$-dependent \gls*{epc} to be studied.

In conclusion, by employing \gls*{rixs} at the O K-edge, we extracted the \gls*{epc} in an ultrathin \gls*{SL} series of the form $n$SrIrO$_3$/$m$\gls*{sto}. Multiple phonon excitations corresponding to the \gls*{lo4} branch are observed for the entire set of \gls*{SL}s, along with a \gls*{sto} substrate. Using a closed form solution for the \gls*{rixs} cross-section, the \gls*{epc} was found to be strongly reduced for samples with $n\geq m$.  With the proposed strong link of Ti-based superconductivity and the \gls*{epc} of the \gls*{lo4} phonon mode, these results highlight heterostructuring as a feasible alternative method to modify \gls{epc} without invoking carrier doping,  providing insight into the superconducting state and its dependence on the \gls*{epc}. Furthermore, these results showcase \gls*{rixs} as the sole feasible method currently able to directly extract the \gls*{epc} in thin film heterostructures, where such measurements are poised to bring deeper insights to how lattice interactions drive changes in electronic properties.

\section*{Acknowledgements}
The authors acknowledge useful discussions with Wei-Guo Yin, Yilin Wang, Lukas Horak, Chris Rouleau, and Neil J.\ Robinson. The authors also acknowledge helpful correspondence with Simon Moser. The authors also acknowledge experimental assisstance from Milan Radovic for use of a SrTiO$_3$ substrate.  This material is based upon work supported by the U.S. Department of Energy, Office of Basic Energy Sciences, Early Career Award Program under Award No. 1047478. Work at Brookhaven National Laboratory was supported by the U.S. Department of Energy, Office of Science, Office of Basic Energy Sciences, under Contract No.~DE-SC0012704. The RIXS experiments were performed at the ADRESS beamline of the Swiss Light Source at the Paul Scherrer Institut. Work at the Paul Scherrer Institut was supported by the Swiss National Science Foundation through the NCCR MARVEL, the SINERGIA network ``Mott Physics beyond the Heisenberg Model" and a D-A-CH project (SNSF Research Grant No. 200021L 141325). J.P. and T.S. acknowledge financial support through the Dysenos AG by Kabelwerke Brugg AG Holding, Fachhochschule Nordwestschweiz, and the Paul Scherrer Institut. J. P. also acknowledges financial support by the Swiss National Science Foundation Early Postdoc. Mobility fellowship project number P2FRP2$\_$171824.  J. L. acknowledges the support by the Science Alliance Joint Directed Research \& Development Program and the Organized Research Unit at the University of Tennessee. J. L. also acknowledges support by the DOD-DARPA under Grant No. HR0011-16-1-0005. A portion of the fabrication, characterization, and theoretical calculations by TB was conducted at the Center for Nanophase Materials Sciences, which is a DOE Office of Science User Facility. This research used resources of the National Energy Research Scientific Computing Center (NERSC), a U.S. Department of Energy Office of Science User Facility operated under Contract No. DE-AC02-05CH11231. Use of the Advanced Photon Source, an Office of Science User Facility operated for the U.S. DOE, OS by Argonne National Laboratory, was supported by the U.S. DOE under Contract No. DE-AC02-06CH11357. This research used resources of the Advanced Light Source, which is a DOE Office of Science User Facility under contract no. DE-AC02-05CH11231.


\begin{thebibliography}{55}%
\makeatletter
\providecommand \@ifxundefined [1]{%
 \@ifx{#1\undefined}
}%
\providecommand \@ifnum [1]{%
 \ifnum #1\expandafter \@firstoftwo
 \else \expandafter \@secondoftwo
 \fi
}%
\providecommand \@ifx [1]{%
 \ifx #1\expandafter \@firstoftwo
 \else \expandafter \@secondoftwo
 \fi
}%
\providecommand \natexlab [1]{#1}%
\providecommand \enquote  [1]{``#1''}%
\providecommand \bibnamefont  [1]{#1}%
\providecommand \bibfnamefont [1]{#1}%
\providecommand \citenamefont [1]{#1}%
\providecommand \href@noop [0]{\@secondoftwo}%
\providecommand \href [0]{\begingroup \@sanitize@url \@href}%
\providecommand \@href[1]{\@@startlink{#1}\@@href}%
\providecommand \@@href[1]{\endgroup#1\@@endlink}%
\providecommand \@sanitize@url [0]{\catcode `\\12\catcode `\$12\catcode
  `\&12\catcode `\#12\catcode `\^12\catcode `\_12\catcode `\%12\relax}%
\providecommand \@@startlink[1]{}%
\providecommand \@@endlink[0]{}%
\providecommand \url  [0]{\begingroup\@sanitize@url \@url }%
\providecommand \@url [1]{\endgroup\@href {#1}{\urlprefix }}%
\providecommand \urlprefix  [0]{URL }%
\providecommand \Eprint [0]{\href }%
\providecommand \doibase [0]{http://dx.doi.org/}%
\providecommand \selectlanguage [0]{\@gobble}%
\providecommand \bibinfo  [0]{\@secondoftwo}%
\providecommand \bibfield  [0]{\@secondoftwo}%
\providecommand \translation [1]{[#1]}%
\providecommand \BibitemOpen [0]{}%
\providecommand \bibitemStop [0]{}%
\providecommand \bibitemNoStop [0]{.\EOS\space}%
\providecommand \EOS [0]{\spacefactor3000\relax}%
\providecommand \BibitemShut  [1]{\csname bibitem#1\endcsname}%
\let\auto@bib@innerbib\@empty
\bibitem [{\citenamefont {Schooley}\ \emph {et~al.}(1964)\citenamefont
  {Schooley}, \citenamefont {Hosler},\ and\ \citenamefont
  {Cohen}}]{Schooley_STO_SC_1964}%
  \BibitemOpen
  \bibfield  {author} {\bibinfo {author} {\bibfnamefont {J.~F.}\ \bibnamefont
  {Schooley}}, \bibinfo {author} {\bibfnamefont {W.~R.}\ \bibnamefont
  {Hosler}}, \ and\ \bibinfo {author} {\bibfnamefont {M.~L.}\ \bibnamefont
  {Cohen}},\ }\href {\doibase 10.1103/PhysRevLett.12.474} {\bibfield  {journal}
  {\bibinfo  {journal} {Phys. Rev. Lett.}\ }\textbf {\bibinfo {volume} {12}},\
  \bibinfo {pages} {474} (\bibinfo {year} {1964})}\BibitemShut {NoStop}%
\bibitem [{\citenamefont {Binnig}\ \emph {et~al.}(1980)\citenamefont {Binnig},
  \citenamefont {Baratoff}, \citenamefont {Hoenig},\ and\ \citenamefont
  {Bednorz}}]{binnig1980two}%
  \BibitemOpen
  \bibfield  {author} {\bibinfo {author} {\bibfnamefont {G.}~\bibnamefont
  {Binnig}}, \bibinfo {author} {\bibfnamefont {A.}~\bibnamefont {Baratoff}},
  \bibinfo {author} {\bibfnamefont {H.}~\bibnamefont {Hoenig}}, \ and\ \bibinfo
  {author} {\bibfnamefont {J.}~\bibnamefont {Bednorz}},\ }\href@noop {}
  {\bibfield  {journal} {\bibinfo  {journal} {Physical Review Letters}\
  }\textbf {\bibinfo {volume} {45}},\ \bibinfo {pages} {1352} (\bibinfo {year}
  {1980})}\BibitemShut {NoStop}%
\bibitem [{\citenamefont {Ohtomo}\ and\ \citenamefont
  {Hwang}(2004)}]{ohtomo2004high}%
  \BibitemOpen
  \bibfield  {author} {\bibinfo {author} {\bibfnamefont {A.}~\bibnamefont
  {Ohtomo}}\ and\ \bibinfo {author} {\bibfnamefont {H.}~\bibnamefont {Hwang}},\
  }\href@noop {} {\bibfield  {journal} {\bibinfo  {journal} {Nature}\ }\textbf
  {\bibinfo {volume} {427}},\ \bibinfo {pages} {423} (\bibinfo {year}
  {2004})}\BibitemShut {NoStop}%
\bibitem [{\citenamefont {Reyren}\ \emph {et~al.}(2007)\citenamefont {Reyren},
  \citenamefont {Thiel}, \citenamefont {Caviglia}, \citenamefont {Kourkoutis},
  \citenamefont {Hammerl}, \citenamefont {Richter}, \citenamefont {Schneider},
  \citenamefont {Kopp}, \citenamefont {R{\"u}etschi}, \citenamefont {Jaccard}
  \emph {et~al.}}]{reyren2007superconducting}%
  \BibitemOpen
  \bibfield  {author} {\bibinfo {author} {\bibfnamefont {N.}~\bibnamefont
  {Reyren}}, \bibinfo {author} {\bibfnamefont {S.}~\bibnamefont {Thiel}},
  \bibinfo {author} {\bibfnamefont {A.}~\bibnamefont {Caviglia}}, \bibinfo
  {author} {\bibfnamefont {L.~F.}\ \bibnamefont {Kourkoutis}}, \bibinfo
  {author} {\bibfnamefont {G.}~\bibnamefont {Hammerl}}, \bibinfo {author}
  {\bibfnamefont {C.}~\bibnamefont {Richter}}, \bibinfo {author} {\bibfnamefont
  {C.}~\bibnamefont {Schneider}}, \bibinfo {author} {\bibfnamefont
  {T.}~\bibnamefont {Kopp}}, \bibinfo {author} {\bibfnamefont {A.-S.}\
  \bibnamefont {R{\"u}etschi}}, \bibinfo {author} {\bibfnamefont
  {D.}~\bibnamefont {Jaccard}},  \emph {et~al.},\ }\href@noop {} {\bibfield
  {journal} {\bibinfo  {journal} {Science}\ }\textbf {\bibinfo {volume}
  {317}},\ \bibinfo {pages} {1196} (\bibinfo {year} {2007})}\BibitemShut
  {NoStop}%
\bibitem [{\citenamefont {Qing-Yan}\ \emph {et~al.}(2012)\citenamefont
  {Qing-Yan}, \citenamefont {Zhi}, \citenamefont {Wen-Hao}, \citenamefont
  {Zuo-Cheng}, \citenamefont {Jin-Song}, \citenamefont {Wei}, \citenamefont
  {Hao}, \citenamefont {Yun-Bo}, \citenamefont {Peng}, \citenamefont {Kai}
  \emph {et~al.}}]{qing2012interface}%
  \BibitemOpen
  \bibfield  {author} {\bibinfo {author} {\bibfnamefont {W.}~\bibnamefont
  {Qing-Yan}}, \bibinfo {author} {\bibfnamefont {L.}~\bibnamefont {Zhi}},
  \bibinfo {author} {\bibfnamefont {Z.}~\bibnamefont {Wen-Hao}}, \bibinfo
  {author} {\bibfnamefont {Z.}~\bibnamefont {Zuo-Cheng}}, \bibinfo {author}
  {\bibfnamefont {Z.}~\bibnamefont {Jin-Song}}, \bibinfo {author}
  {\bibfnamefont {L.}~\bibnamefont {Wei}}, \bibinfo {author} {\bibfnamefont
  {D.}~\bibnamefont {Hao}}, \bibinfo {author} {\bibfnamefont {O.}~\bibnamefont
  {Yun-Bo}}, \bibinfo {author} {\bibfnamefont {D.}~\bibnamefont {Peng}},
  \bibinfo {author} {\bibfnamefont {C.}~\bibnamefont {Kai}},  \emph {et~al.},\
  }\href@noop {} {\bibfield  {journal} {\bibinfo  {journal} {Chinese Physics
  Letters}\ }\textbf {\bibinfo {volume} {29}},\ \bibinfo {pages} {037402}
  (\bibinfo {year} {2012})}\BibitemShut {NoStop}%
\bibitem [{\citenamefont {Liu}\ \emph {et~al.}(2012)\citenamefont {Liu},
  \citenamefont {Zhang}, \citenamefont {Mou}, \citenamefont {He}, \citenamefont
  {Ou}, \citenamefont {Wang}, \citenamefont {Li}, \citenamefont {Wang},
  \citenamefont {Zhao}, \citenamefont {He} \emph {et~al.}}]{liu2012electronic}%
  \BibitemOpen
  \bibfield  {author} {\bibinfo {author} {\bibfnamefont {D.}~\bibnamefont
  {Liu}}, \bibinfo {author} {\bibfnamefont {W.}~\bibnamefont {Zhang}}, \bibinfo
  {author} {\bibfnamefont {D.}~\bibnamefont {Mou}}, \bibinfo {author}
  {\bibfnamefont {J.}~\bibnamefont {He}}, \bibinfo {author} {\bibfnamefont
  {Y.-B.}\ \bibnamefont {Ou}}, \bibinfo {author} {\bibfnamefont {Q.-Y.}\
  \bibnamefont {Wang}}, \bibinfo {author} {\bibfnamefont {Z.}~\bibnamefont
  {Li}}, \bibinfo {author} {\bibfnamefont {L.}~\bibnamefont {Wang}}, \bibinfo
  {author} {\bibfnamefont {L.}~\bibnamefont {Zhao}}, \bibinfo {author}
  {\bibfnamefont {S.}~\bibnamefont {He}},  \emph {et~al.},\ }\href@noop {}
  {\bibfield  {journal} {\bibinfo  {journal} {arXiv preprint arXiv:1202.5849}\
  } (\bibinfo {year} {2012})}\BibitemShut {NoStop}%
\bibitem [{\citenamefont {He}\ \emph {et~al.}(2013)\citenamefont {He},
  \citenamefont {He}, \citenamefont {Zhang}, \citenamefont {Zhao},
  \citenamefont {Liu}, \citenamefont {Liu}, \citenamefont {Mou}, \citenamefont
  {Ou}, \citenamefont {Wang}, \citenamefont {Li} \emph {et~al.}}]{he2013phase}%
  \BibitemOpen
  \bibfield  {author} {\bibinfo {author} {\bibfnamefont {S.}~\bibnamefont
  {He}}, \bibinfo {author} {\bibfnamefont {J.}~\bibnamefont {He}}, \bibinfo
  {author} {\bibfnamefont {W.}~\bibnamefont {Zhang}}, \bibinfo {author}
  {\bibfnamefont {L.}~\bibnamefont {Zhao}}, \bibinfo {author} {\bibfnamefont
  {D.}~\bibnamefont {Liu}}, \bibinfo {author} {\bibfnamefont {X.}~\bibnamefont
  {Liu}}, \bibinfo {author} {\bibfnamefont {D.}~\bibnamefont {Mou}}, \bibinfo
  {author} {\bibfnamefont {Y.-B.}\ \bibnamefont {Ou}}, \bibinfo {author}
  {\bibfnamefont {Q.-Y.}\ \bibnamefont {Wang}}, \bibinfo {author}
  {\bibfnamefont {Z.}~\bibnamefont {Li}},  \emph {et~al.},\ }\href@noop {}
  {\bibfield  {journal} {\bibinfo  {journal} {Nature materials}\ }\textbf
  {\bibinfo {volume} {12}},\ \bibinfo {pages} {605} (\bibinfo {year}
  {2013})}\BibitemShut {NoStop}%
\bibitem [{\citenamefont {Cowley}(1964)}]{cowley1964lattice}%
  \BibitemOpen
  \bibfield  {author} {\bibinfo {author} {\bibfnamefont {R.}~\bibnamefont
  {Cowley}},\ }\href@noop {} {\bibfield  {journal} {\bibinfo  {journal}
  {Physical Review}\ }\textbf {\bibinfo {volume} {134}},\ \bibinfo {pages}
  {A981} (\bibinfo {year} {1964})}\BibitemShut {NoStop}%
\bibitem [{\citenamefont {Wang}\ \emph {et~al.}(2015)\citenamefont {Wang},
  \citenamefont {Walker}, \citenamefont {Tamai}, \citenamefont {Wang},
  \citenamefont {Ristic}, \citenamefont {Bruno}, \citenamefont {De~La~Torre},
  \citenamefont {Ricc{\`o}}, \citenamefont {Plumb}, \citenamefont {Shi} \emph
  {et~al.}}]{wang2015tailoring}%
  \BibitemOpen
  \bibfield  {author} {\bibinfo {author} {\bibfnamefont {Z.}~\bibnamefont
  {Wang}}, \bibinfo {author} {\bibfnamefont {S.~M.}\ \bibnamefont {Walker}},
  \bibinfo {author} {\bibfnamefont {A.}~\bibnamefont {Tamai}}, \bibinfo
  {author} {\bibfnamefont {Y.}~\bibnamefont {Wang}}, \bibinfo {author}
  {\bibfnamefont {Z.}~\bibnamefont {Ristic}}, \bibinfo {author} {\bibfnamefont
  {F.~Y.}\ \bibnamefont {Bruno}}, \bibinfo {author} {\bibfnamefont
  {A.}~\bibnamefont {De~La~Torre}}, \bibinfo {author} {\bibfnamefont
  {S.}~\bibnamefont {Ricc{\`o}}}, \bibinfo {author} {\bibfnamefont
  {N.}~\bibnamefont {Plumb}}, \bibinfo {author} {\bibfnamefont
  {M.}~\bibnamefont {Shi}},  \emph {et~al.},\ }\href@noop {} {\bibfield
  {journal} {\bibinfo  {journal} {Nature Materias}\ }\textbf {\bibinfo {volume}
  {15}},\ \bibinfo {pages} {835} (\bibinfo {year} {2015})}\BibitemShut
  {NoStop}%
\bibitem [{\citenamefont {Gervais}\ \emph {et~al.}(1993)\citenamefont
  {Gervais}, \citenamefont {Servoin}, \citenamefont {Baratoff}, \citenamefont
  {Bednorz},\ and\ \citenamefont {Binnig}}]{gervais1993temperature}%
  \BibitemOpen
  \bibfield  {author} {\bibinfo {author} {\bibfnamefont {F.}~\bibnamefont
  {Gervais}}, \bibinfo {author} {\bibfnamefont {J.-L.}\ \bibnamefont
  {Servoin}}, \bibinfo {author} {\bibfnamefont {A.}~\bibnamefont {Baratoff}},
  \bibinfo {author} {\bibfnamefont {J.~G.}\ \bibnamefont {Bednorz}}, \ and\
  \bibinfo {author} {\bibfnamefont {G.}~\bibnamefont {Binnig}},\ }\href@noop {}
  {\bibfield  {journal} {\bibinfo  {journal} {Physical Review B}\ }\textbf
  {\bibinfo {volume} {47}},\ \bibinfo {pages} {8187} (\bibinfo {year}
  {1993})}\BibitemShut {NoStop}%
\bibitem [{\citenamefont {Rosenstein}\ \emph {et~al.}(2016)\citenamefont
  {Rosenstein}, \citenamefont {Shapiro}, \citenamefont {Shapiro},\ and\
  \citenamefont {Li}}]{rosenstein2016superconductivity}%
  \BibitemOpen
  \bibfield  {author} {\bibinfo {author} {\bibfnamefont {B.}~\bibnamefont
  {Rosenstein}}, \bibinfo {author} {\bibfnamefont {B.~Y.}\ \bibnamefont
  {Shapiro}}, \bibinfo {author} {\bibfnamefont {I.}~\bibnamefont {Shapiro}}, \
  and\ \bibinfo {author} {\bibfnamefont {D.}~\bibnamefont {Li}},\ }\href@noop
  {} {\bibfield  {journal} {\bibinfo  {journal} {Physical Review B}\ }\textbf
  {\bibinfo {volume} {94}},\ \bibinfo {pages} {024505} (\bibinfo {year}
  {2016})}\BibitemShut {NoStop}%
\bibitem [{\citenamefont {Baratoff}\ and\ \citenamefont
  {Binnig}(1981)}]{baratoff1981mechanism}%
  \BibitemOpen
  \bibfield  {author} {\bibinfo {author} {\bibfnamefont {A.}~\bibnamefont
  {Baratoff}}\ and\ \bibinfo {author} {\bibfnamefont {G.}~\bibnamefont
  {Binnig}},\ }\href@noop {} {\bibfield  {journal} {\bibinfo  {journal}
  {Physica B+ C}\ }\textbf {\bibinfo {volume} {108}},\ \bibinfo {pages} {1335}
  (\bibinfo {year} {1981})}\BibitemShut {NoStop}%
\bibitem [{\citenamefont {Swartz}\ \emph {et~al.}(2018)\citenamefont {Swartz},
  \citenamefont {Inoue}, \citenamefont {Merz}, \citenamefont {Hikita},
  \citenamefont {Raghu}, \citenamefont {Devereaux}, \citenamefont {Johnston},\
  and\ \citenamefont {Hwang}}]{Swartz2018}%
  \BibitemOpen
  \bibfield  {author} {\bibinfo {author} {\bibfnamefont {A.~G.}\ \bibnamefont
  {Swartz}}, \bibinfo {author} {\bibfnamefont {H.}~\bibnamefont {Inoue}},
  \bibinfo {author} {\bibfnamefont {T.~A.}\ \bibnamefont {Merz}}, \bibinfo
  {author} {\bibfnamefont {Y.}~\bibnamefont {Hikita}}, \bibinfo {author}
  {\bibfnamefont {S.}~\bibnamefont {Raghu}}, \bibinfo {author} {\bibfnamefont
  {T.~P.}\ \bibnamefont {Devereaux}}, \bibinfo {author} {\bibfnamefont
  {S.}~\bibnamefont {Johnston}}, \ and\ \bibinfo {author} {\bibfnamefont
  {H.~Y.}\ \bibnamefont {Hwang}},\ }\href {\doibase 10.1073/pnas.1713916115}
  {\bibfield  {journal} {\bibinfo  {journal} {Proceedings of the National
  Academy of Sciences}\ }\textbf {\bibinfo {volume} {115}},\ \bibinfo {pages}
  {1475} (\bibinfo {year} {2018})},\ \Eprint
  {http://arxiv.org/abs/http://www.pnas.org/content/115/7/1475.full.pdf}
  {http://www.pnas.org/content/115/7/1475.full.pdf} \BibitemShut {NoStop}%
\bibitem [{\citenamefont {Moser}\ \emph {et~al.}(2013)\citenamefont {Moser},
  \citenamefont {Moreschini}, \citenamefont {Ja{\'c}imovi{\'c}}, \citenamefont
  {Bari{\v{s}}i{\'c}}, \citenamefont {Berger}, \citenamefont {Magrez},
  \citenamefont {Chang}, \citenamefont {Kim}, \citenamefont {Bostwick},
  \citenamefont {Rotenberg} \emph {et~al.}}]{moser2013tunable}%
  \BibitemOpen
  \bibfield  {author} {\bibinfo {author} {\bibfnamefont {S.}~\bibnamefont
  {Moser}}, \bibinfo {author} {\bibfnamefont {L.}~\bibnamefont {Moreschini}},
  \bibinfo {author} {\bibfnamefont {J.}~\bibnamefont {Ja{\'c}imovi{\'c}}},
  \bibinfo {author} {\bibfnamefont {O.}~\bibnamefont {Bari{\v{s}}i{\'c}}},
  \bibinfo {author} {\bibfnamefont {H.}~\bibnamefont {Berger}}, \bibinfo
  {author} {\bibfnamefont {A.}~\bibnamefont {Magrez}}, \bibinfo {author}
  {\bibfnamefont {Y.}~\bibnamefont {Chang}}, \bibinfo {author} {\bibfnamefont
  {K.}~\bibnamefont {Kim}}, \bibinfo {author} {\bibfnamefont {A.}~\bibnamefont
  {Bostwick}}, \bibinfo {author} {\bibfnamefont {E.}~\bibnamefont {Rotenberg}},
   \emph {et~al.},\ }\href@noop {} {\bibfield  {journal} {\bibinfo  {journal}
  {Physical review letters}\ }\textbf {\bibinfo {volume} {110}},\ \bibinfo
  {pages} {196403} (\bibinfo {year} {2013})}\BibitemShut {NoStop}%
\bibitem [{\citenamefont
  {Giustino}(2017{\natexlab{a}})}]{giustino2017electron}%
  \BibitemOpen
  \bibfield  {author} {\bibinfo {author} {\bibfnamefont {F.}~\bibnamefont
  {Giustino}},\ }\href@noop {} {\bibfield  {journal} {\bibinfo  {journal}
  {Reviews of Modern Physics}\ }\textbf {\bibinfo {volume} {89}},\ \bibinfo
  {pages} {015003} (\bibinfo {year} {2017}{\natexlab{a}})}\BibitemShut
  {NoStop}%
\bibitem [{\citenamefont {Zhou}\ \emph {et~al.}(2011)\citenamefont {Zhou},
  \citenamefont {Radovic}, \citenamefont {Schlappa}, \citenamefont {Strocov},
  \citenamefont {Frison}, \citenamefont {Mesot}, \citenamefont {Patthey},\ and\
  \citenamefont {Schmitt}}]{zhou2011localized}%
  \BibitemOpen
  \bibfield  {author} {\bibinfo {author} {\bibfnamefont {K.-J.}\ \bibnamefont
  {Zhou}}, \bibinfo {author} {\bibfnamefont {M.}~\bibnamefont {Radovic}},
  \bibinfo {author} {\bibfnamefont {J.}~\bibnamefont {Schlappa}}, \bibinfo
  {author} {\bibfnamefont {V.}~\bibnamefont {Strocov}}, \bibinfo {author}
  {\bibfnamefont {R.}~\bibnamefont {Frison}}, \bibinfo {author} {\bibfnamefont
  {J.}~\bibnamefont {Mesot}}, \bibinfo {author} {\bibfnamefont
  {L.}~\bibnamefont {Patthey}}, \ and\ \bibinfo {author} {\bibfnamefont
  {T.}~\bibnamefont {Schmitt}},\ }\href@noop {} {\bibfield  {journal} {\bibinfo
   {journal} {Physical Review B}\ }\textbf {\bibinfo {volume} {83}},\ \bibinfo
  {pages} {201402} (\bibinfo {year} {2011})}\BibitemShut {NoStop}%
\bibitem [{\citenamefont {Fatale}\ \emph {et~al.}(2016)\citenamefont {Fatale},
  \citenamefont {Moser}, \citenamefont {Miyawaki}, \citenamefont {Harada},\
  and\ \citenamefont {Grioni}}]{fatale2016hybridization}%
  \BibitemOpen
  \bibfield  {author} {\bibinfo {author} {\bibfnamefont {S.}~\bibnamefont
  {Fatale}}, \bibinfo {author} {\bibfnamefont {S.}~\bibnamefont {Moser}},
  \bibinfo {author} {\bibfnamefont {J.}~\bibnamefont {Miyawaki}}, \bibinfo
  {author} {\bibfnamefont {Y.}~\bibnamefont {Harada}}, \ and\ \bibinfo {author}
  {\bibfnamefont {M.}~\bibnamefont {Grioni}},\ }\href@noop {} {\bibfield
  {journal} {\bibinfo  {journal} {Physical Review B}\ }\textbf {\bibinfo
  {volume} {94}},\ \bibinfo {pages} {195131} (\bibinfo {year}
  {2016})}\BibitemShut {NoStop}%
\bibitem [{\citenamefont {Lee}\ \emph {et~al.}(2014)\citenamefont {Lee},
  \citenamefont {Moritz}, \citenamefont {Lee}, \citenamefont {Yi},
  \citenamefont {Jia}, \citenamefont {Sorini}, \citenamefont {Kudo},
  \citenamefont {Koike}, \citenamefont {Zhou}, \citenamefont {Monney} \emph
  {et~al.}}]{lee2014charge}%
  \BibitemOpen
  \bibfield  {author} {\bibinfo {author} {\bibfnamefont {J.}~\bibnamefont
  {Lee}}, \bibinfo {author} {\bibfnamefont {B.}~\bibnamefont {Moritz}},
  \bibinfo {author} {\bibfnamefont {W.}~\bibnamefont {Lee}}, \bibinfo {author}
  {\bibfnamefont {M.}~\bibnamefont {Yi}}, \bibinfo {author} {\bibfnamefont
  {C.}~\bibnamefont {Jia}}, \bibinfo {author} {\bibfnamefont {A.}~\bibnamefont
  {Sorini}}, \bibinfo {author} {\bibfnamefont {K.}~\bibnamefont {Kudo}},
  \bibinfo {author} {\bibfnamefont {Y.}~\bibnamefont {Koike}}, \bibinfo
  {author} {\bibfnamefont {K.}~\bibnamefont {Zhou}}, \bibinfo {author}
  {\bibfnamefont {C.}~\bibnamefont {Monney}},  \emph {et~al.},\ }\href@noop {}
  {\bibfield  {journal} {\bibinfo  {journal} {Physical Review B}\ }\textbf
  {\bibinfo {volume} {89}},\ \bibinfo {pages} {041104} (\bibinfo {year}
  {2014})}\BibitemShut {NoStop}%
\bibitem [{\citenamefont {Johnston}\ \emph {et~al.}(2016)\citenamefont
  {Johnston}, \citenamefont {Monney}, \citenamefont {Bisogni}, \citenamefont
  {Zhou}, \citenamefont {Kraus}, \citenamefont {Behr}, \citenamefont {Strocov},
  \citenamefont {M{\'a}lek}, \citenamefont {Drechsler}, \citenamefont {Geck}
  \emph {et~al.}}]{johnston2016electron}%
  \BibitemOpen
  \bibfield  {author} {\bibinfo {author} {\bibfnamefont {S.}~\bibnamefont
  {Johnston}}, \bibinfo {author} {\bibfnamefont {C.}~\bibnamefont {Monney}},
  \bibinfo {author} {\bibfnamefont {V.}~\bibnamefont {Bisogni}}, \bibinfo
  {author} {\bibfnamefont {K.-J.}\ \bibnamefont {Zhou}}, \bibinfo {author}
  {\bibfnamefont {R.}~\bibnamefont {Kraus}}, \bibinfo {author} {\bibfnamefont
  {G.}~\bibnamefont {Behr}}, \bibinfo {author} {\bibfnamefont {V.~N.}\
  \bibnamefont {Strocov}}, \bibinfo {author} {\bibfnamefont {J.}~\bibnamefont
  {M{\'a}lek}}, \bibinfo {author} {\bibfnamefont {S.-L.}\ \bibnamefont
  {Drechsler}}, \bibinfo {author} {\bibfnamefont {J.}~\bibnamefont {Geck}},
  \emph {et~al.},\ }\href@noop {} {\bibfield  {journal} {\bibinfo  {journal}
  {Nature communications}\ }\textbf {\bibinfo {volume} {7}} (\bibinfo {year}
  {2016})}\BibitemShut {NoStop}%
\bibitem [{\citenamefont {Lee}\ \emph {et~al.}(2013{\natexlab{a}})\citenamefont
  {Lee}, \citenamefont {Johnston}, \citenamefont {Moritz}, \citenamefont {Lee},
  \citenamefont {Yi}, \citenamefont {Zhou}, \citenamefont {Schmitt},
  \citenamefont {Patthey}, \citenamefont {Strocov}, \citenamefont {Kudo} \emph
  {et~al.}}]{lee2013role}%
  \BibitemOpen
  \bibfield  {author} {\bibinfo {author} {\bibfnamefont {W.}~\bibnamefont
  {Lee}}, \bibinfo {author} {\bibfnamefont {S.}~\bibnamefont {Johnston}},
  \bibinfo {author} {\bibfnamefont {B.}~\bibnamefont {Moritz}}, \bibinfo
  {author} {\bibfnamefont {J.}~\bibnamefont {Lee}}, \bibinfo {author}
  {\bibfnamefont {M.}~\bibnamefont {Yi}}, \bibinfo {author} {\bibfnamefont
  {K.}~\bibnamefont {Zhou}}, \bibinfo {author} {\bibfnamefont {T.}~\bibnamefont
  {Schmitt}}, \bibinfo {author} {\bibfnamefont {L.}~\bibnamefont {Patthey}},
  \bibinfo {author} {\bibfnamefont {V.}~\bibnamefont {Strocov}}, \bibinfo
  {author} {\bibfnamefont {K.}~\bibnamefont {Kudo}},  \emph {et~al.},\
  }\href@noop {} {\bibfield  {journal} {\bibinfo  {journal} {Physical review
  letters}\ }\textbf {\bibinfo {volume} {110}},\ \bibinfo {pages} {265502}
  (\bibinfo {year} {2013}{\natexlab{a}})}\BibitemShut {NoStop}%
\bibitem [{\citenamefont {Moser}\ \emph {et~al.}(2015)\citenamefont {Moser},
  \citenamefont {Fatale}, \citenamefont {Kr{\"u}ger}, \citenamefont {Berger},
  \citenamefont {Bugnon}, \citenamefont {Magrez}, \citenamefont {Niwa},
  \citenamefont {Miyawaki}, \citenamefont {Harada},\ and\ \citenamefont
  {Grioni}}]{moser2015electron}%
  \BibitemOpen
  \bibfield  {author} {\bibinfo {author} {\bibfnamefont {S.}~\bibnamefont
  {Moser}}, \bibinfo {author} {\bibfnamefont {S.}~\bibnamefont {Fatale}},
  \bibinfo {author} {\bibfnamefont {P.}~\bibnamefont {Kr{\"u}ger}}, \bibinfo
  {author} {\bibfnamefont {H.}~\bibnamefont {Berger}}, \bibinfo {author}
  {\bibfnamefont {P.}~\bibnamefont {Bugnon}}, \bibinfo {author} {\bibfnamefont
  {A.}~\bibnamefont {Magrez}}, \bibinfo {author} {\bibfnamefont
  {H.}~\bibnamefont {Niwa}}, \bibinfo {author} {\bibfnamefont {J.}~\bibnamefont
  {Miyawaki}}, \bibinfo {author} {\bibfnamefont {Y.}~\bibnamefont {Harada}}, \
  and\ \bibinfo {author} {\bibfnamefont {M.}~\bibnamefont {Grioni}},\
  }\href@noop {} {\bibfield  {journal} {\bibinfo  {journal} {Physical review
  letters}\ }\textbf {\bibinfo {volume} {115}},\ \bibinfo {pages} {096404}
  (\bibinfo {year} {2015})}\BibitemShut {NoStop}%
\bibitem [{\citenamefont {Ament}\ \emph
  {et~al.}(2011{\natexlab{a}})\citenamefont {Ament}, \citenamefont
  {Van~Veenendaal},\ and\ \citenamefont {Van
  Den~Brink}}]{ament2011determining}%
  \BibitemOpen
  \bibfield  {author} {\bibinfo {author} {\bibfnamefont {L.}~\bibnamefont
  {Ament}}, \bibinfo {author} {\bibfnamefont {M.}~\bibnamefont
  {Van~Veenendaal}}, \ and\ \bibinfo {author} {\bibfnamefont {J.}~\bibnamefont
  {Van Den~Brink}},\ }\href@noop {} {\bibfield  {journal} {\bibinfo  {journal}
  {Europhysics Letters}\ }\textbf {\bibinfo {volume} {95}},\ \bibinfo {pages}
  {27008} (\bibinfo {year} {2011}{\natexlab{a}})}\BibitemShut {NoStop}%
\bibitem [{sup()}]{supplemental}%
  \BibitemOpen
  \href@noop {} {}\bibinfo {note} {See Supplemental Materials at [URL] for
  experimental and computational details.}\BibitemShut {Stop}%
\bibitem [{\citenamefont {Lee}\ \emph {et~al.}(2013{\natexlab{b}})\citenamefont
  {Lee}, \citenamefont {Schmitt}, \citenamefont {Moore}, \citenamefont
  {Johnston}, \citenamefont {Cui}, \citenamefont {Li}, \citenamefont {Yi},
  \citenamefont {Liu}, \citenamefont {Hashimoto}, \citenamefont {Zhang} \emph
  {et~al.}}]{lee2013interfacial}%
  \BibitemOpen
  \bibfield  {author} {\bibinfo {author} {\bibfnamefont {J.}~\bibnamefont
  {Lee}}, \bibinfo {author} {\bibfnamefont {F.}~\bibnamefont {Schmitt}},
  \bibinfo {author} {\bibfnamefont {R.}~\bibnamefont {Moore}}, \bibinfo
  {author} {\bibfnamefont {S.}~\bibnamefont {Johnston}}, \bibinfo {author}
  {\bibfnamefont {Y.-T.}\ \bibnamefont {Cui}}, \bibinfo {author} {\bibfnamefont
  {W.}~\bibnamefont {Li}}, \bibinfo {author} {\bibfnamefont {M.}~\bibnamefont
  {Yi}}, \bibinfo {author} {\bibfnamefont {Z.}~\bibnamefont {Liu}}, \bibinfo
  {author} {\bibfnamefont {M.}~\bibnamefont {Hashimoto}}, \bibinfo {author}
  {\bibfnamefont {Y.}~\bibnamefont {Zhang}},  \emph {et~al.},\ }\href@noop {}
  {\bibfield  {journal} {\bibinfo  {journal} {Nature}\ }\textbf {\bibinfo
  {volume} {515}},\ \bibinfo {pages} {245} (\bibinfo {year}
  {2013}{\natexlab{b}})}\BibitemShut {NoStop}%
\bibitem [{\citenamefont {Devereaux}\ \emph {et~al.}(2016)\citenamefont
  {Devereaux}, \citenamefont {Shvaika}, \citenamefont {Wu}, \citenamefont
  {Wohlfeld}, \citenamefont {Jia}, \citenamefont {Wang}, \citenamefont
  {Moritz}, \citenamefont {Chaix}, \citenamefont {Lee}, \citenamefont {Shen}
  \emph {et~al.}}]{devereaux2016directly}%
  \BibitemOpen
  \bibfield  {author} {\bibinfo {author} {\bibfnamefont {T.}~\bibnamefont
  {Devereaux}}, \bibinfo {author} {\bibfnamefont {A.}~\bibnamefont {Shvaika}},
  \bibinfo {author} {\bibfnamefont {K.}~\bibnamefont {Wu}}, \bibinfo {author}
  {\bibfnamefont {K.}~\bibnamefont {Wohlfeld}}, \bibinfo {author}
  {\bibfnamefont {C.}~\bibnamefont {Jia}}, \bibinfo {author} {\bibfnamefont
  {Y.}~\bibnamefont {Wang}}, \bibinfo {author} {\bibfnamefont {B.}~\bibnamefont
  {Moritz}}, \bibinfo {author} {\bibfnamefont {L.}~\bibnamefont {Chaix}},
  \bibinfo {author} {\bibfnamefont {W.-S.}\ \bibnamefont {Lee}}, \bibinfo
  {author} {\bibfnamefont {Z.-X.}\ \bibnamefont {Shen}},  \emph {et~al.},\
  }\href@noop {} {\bibfield  {journal} {\bibinfo  {journal} {Physical Review
  X}\ }\textbf {\bibinfo {volume} {6}},\ \bibinfo {pages} {041019} (\bibinfo
  {year} {2016})}\BibitemShut {NoStop}%
\bibitem [{\citenamefont {Hao}\ \emph {et~al.}(2017)\citenamefont {Hao},
  \citenamefont {Meyers}, \citenamefont {Frederick}, \citenamefont {Fabbris},
  \citenamefont {Yang}, \citenamefont {Traynor}, \citenamefont {Horak},
  \citenamefont {Kriegner}, \citenamefont {Choi}, \citenamefont {Kim},
  \citenamefont {Haskel}, \citenamefont {Ryan}, \citenamefont {Dean},\ and\
  \citenamefont {Liu}}]{Hao_arxiv}%
  \BibitemOpen
  \bibfield  {author} {\bibinfo {author} {\bibfnamefont {L.}~\bibnamefont
  {Hao}}, \bibinfo {author} {\bibfnamefont {D.}~\bibnamefont {Meyers}},
  \bibinfo {author} {\bibfnamefont {C.}~\bibnamefont {Frederick}}, \bibinfo
  {author} {\bibfnamefont {G.}~\bibnamefont {Fabbris}}, \bibinfo {author}
  {\bibfnamefont {J.}~\bibnamefont {Yang}}, \bibinfo {author} {\bibfnamefont
  {N.}~\bibnamefont {Traynor}}, \bibinfo {author} {\bibfnamefont
  {L.}~\bibnamefont {Horak}}, \bibinfo {author} {\bibfnamefont
  {D.}~\bibnamefont {Kriegner}}, \bibinfo {author} {\bibfnamefont
  {Y.}~\bibnamefont {Choi}}, \bibinfo {author} {\bibfnamefont {J.-W.}\
  \bibnamefont {Kim}}, \bibinfo {author} {\bibfnamefont {D.}~\bibnamefont
  {Haskel}}, \bibinfo {author} {\bibfnamefont {P.~J.}\ \bibnamefont {Ryan}},
  \bibinfo {author} {\bibfnamefont {M.~P.~M.}\ \bibnamefont {Dean}}, \ and\
  \bibinfo {author} {\bibfnamefont {J.}~\bibnamefont {Liu}},\ }\href {\doibase
  10.1103/PhysRevLett.119.027204} {\bibfield  {journal} {\bibinfo  {journal}
  {Phys. Rev. Lett.}\ }\textbf {\bibinfo {volume} {119}},\ \bibinfo {pages}
  {027204} (\bibinfo {year} {2017})}\BibitemShut {NoStop}%
\bibitem [{\citenamefont {Ghiringhelli}\ \emph {et~al.}(2006)\citenamefont
  {Ghiringhelli}, \citenamefont {Piazzalunga}, \citenamefont {Dallera},
  \citenamefont {Trezzi}, \citenamefont {Braicovich}, \citenamefont {Schmitt},
  \citenamefont {Strocov}, \citenamefont {Betemps}, \citenamefont {Patthey},
  \citenamefont {Wang} \emph {et~al.}}]{ghiringhelli2006saxes}%
  \BibitemOpen
  \bibfield  {author} {\bibinfo {author} {\bibfnamefont {G.}~\bibnamefont
  {Ghiringhelli}}, \bibinfo {author} {\bibfnamefont {A.}~\bibnamefont
  {Piazzalunga}}, \bibinfo {author} {\bibfnamefont {C.}~\bibnamefont
  {Dallera}}, \bibinfo {author} {\bibfnamefont {G.}~\bibnamefont {Trezzi}},
  \bibinfo {author} {\bibfnamefont {L.}~\bibnamefont {Braicovich}}, \bibinfo
  {author} {\bibfnamefont {T.}~\bibnamefont {Schmitt}}, \bibinfo {author}
  {\bibfnamefont {V.}~\bibnamefont {Strocov}}, \bibinfo {author} {\bibfnamefont
  {R.}~\bibnamefont {Betemps}}, \bibinfo {author} {\bibfnamefont
  {L.}~\bibnamefont {Patthey}}, \bibinfo {author} {\bibfnamefont
  {X.}~\bibnamefont {Wang}},  \emph {et~al.},\ }\href@noop {} {\bibfield
  {journal} {\bibinfo  {journal} {Review of Scientific Instruments}\ }\textbf
  {\bibinfo {volume} {77}},\ \bibinfo {pages} {113108} (\bibinfo {year}
  {2006})}\BibitemShut {NoStop}%
\bibitem [{\citenamefont {Schmitt}\ \emph {et~al.}(2013)\citenamefont
  {Schmitt}, \citenamefont {Strocov}, \citenamefont {Zhou}, \citenamefont
  {Schlappa}, \citenamefont {Monney}, \citenamefont {Flechsig},\ and\
  \citenamefont {Patthey}}]{schmitt2013high}%
  \BibitemOpen
  \bibfield  {author} {\bibinfo {author} {\bibfnamefont {T.}~\bibnamefont
  {Schmitt}}, \bibinfo {author} {\bibfnamefont {V.~N.}\ \bibnamefont
  {Strocov}}, \bibinfo {author} {\bibfnamefont {K.-J.}\ \bibnamefont {Zhou}},
  \bibinfo {author} {\bibfnamefont {J.}~\bibnamefont {Schlappa}}, \bibinfo
  {author} {\bibfnamefont {C.}~\bibnamefont {Monney}}, \bibinfo {author}
  {\bibfnamefont {U.}~\bibnamefont {Flechsig}}, \ and\ \bibinfo {author}
  {\bibfnamefont {L.}~\bibnamefont {Patthey}},\ }\href@noop {} {\bibfield
  {journal} {\bibinfo  {journal} {Journal of Electron Spectroscopy and Related
  Phenomena}\ }\textbf {\bibinfo {volume} {188}},\ \bibinfo {pages} {38}
  (\bibinfo {year} {2013})}\BibitemShut {NoStop}%
\bibitem [{Note1()}]{Note1}%
  \BibitemOpen
  \bibinfo {note} {Here r. l. u. is defined within the structural Brillouin
  Zone with $a$ = $b$ = 3.905, $c$ $\approx $ 3.965.}\BibitemShut {Stop}%
\bibitem [{\citenamefont {Serrao}\ \emph {et~al.}(2013)\citenamefont {Serrao},
  \citenamefont {Liu}, \citenamefont {Heron}, \citenamefont {Singh-Bhalla},
  \citenamefont {Yadav}, \citenamefont {Suresha}, \citenamefont {Paull},
  \citenamefont {Yi}, \citenamefont {Chu}, \citenamefont {Trassin} \emph
  {et~al.}}]{serrao2013epitaxy}%
  \BibitemOpen
  \bibfield  {author} {\bibinfo {author} {\bibfnamefont {C.~R.}\ \bibnamefont
  {Serrao}}, \bibinfo {author} {\bibfnamefont {J.}~\bibnamefont {Liu}},
  \bibinfo {author} {\bibfnamefont {J.}~\bibnamefont {Heron}}, \bibinfo
  {author} {\bibfnamefont {G.}~\bibnamefont {Singh-Bhalla}}, \bibinfo {author}
  {\bibfnamefont {A.}~\bibnamefont {Yadav}}, \bibinfo {author} {\bibfnamefont
  {S.}~\bibnamefont {Suresha}}, \bibinfo {author} {\bibfnamefont
  {R.}~\bibnamefont {Paull}}, \bibinfo {author} {\bibfnamefont
  {D.}~\bibnamefont {Yi}}, \bibinfo {author} {\bibfnamefont {J.-H.}\
  \bibnamefont {Chu}}, \bibinfo {author} {\bibfnamefont {M.}~\bibnamefont
  {Trassin}},  \emph {et~al.},\ }\href@noop {} {\bibfield  {journal} {\bibinfo
  {journal} {Physical Review B}\ }\textbf {\bibinfo {volume} {87}},\ \bibinfo
  {pages} {085121} (\bibinfo {year} {2013})}\BibitemShut {NoStop}%
\bibitem [{\citenamefont {de~Groot}\ \emph {et~al.}(1989)\citenamefont
  {de~Groot}, \citenamefont {Grioni}, \citenamefont {Fuggle}, \citenamefont
  {Ghijsen}, \citenamefont {Sawatzky},\ and\ \citenamefont
  {Petersen}}]{Groot_OKedge_2017}%
  \BibitemOpen
  \bibfield  {author} {\bibinfo {author} {\bibfnamefont {F.~M.~F.}\
  \bibnamefont {de~Groot}}, \bibinfo {author} {\bibfnamefont {M.}~\bibnamefont
  {Grioni}}, \bibinfo {author} {\bibfnamefont {J.~C.}\ \bibnamefont {Fuggle}},
  \bibinfo {author} {\bibfnamefont {J.}~\bibnamefont {Ghijsen}}, \bibinfo
  {author} {\bibfnamefont {G.~A.}\ \bibnamefont {Sawatzky}}, \ and\ \bibinfo
  {author} {\bibfnamefont {H.}~\bibnamefont {Petersen}},\ }\href {\doibase
  10.1103/PhysRevB.40.5715} {\bibfield  {journal} {\bibinfo  {journal} {Phys.
  Rev. B}\ }\textbf {\bibinfo {volume} {40}},\ \bibinfo {pages} {5715}
  (\bibinfo {year} {1989})}\BibitemShut {NoStop}%
\bibitem [{\citenamefont {Cao}\ \emph {et~al.}(2016{\natexlab{a}})\citenamefont
  {Cao}, \citenamefont {Park}, \citenamefont {Liu}, \citenamefont {Choudhury},
  \citenamefont {Middey}, \citenamefont {Meyers}, \citenamefont {Kareev},
  \citenamefont {Shafer}, \citenamefont {Arenholz},\ and\ \citenamefont
  {Chakhalian}}]{cao2016orbital}%
  \BibitemOpen
  \bibfield  {author} {\bibinfo {author} {\bibfnamefont {Y.}~\bibnamefont
  {Cao}}, \bibinfo {author} {\bibfnamefont {S.~Y.}\ \bibnamefont {Park}},
  \bibinfo {author} {\bibfnamefont {X.}~\bibnamefont {Liu}}, \bibinfo {author}
  {\bibfnamefont {D.}~\bibnamefont {Choudhury}}, \bibinfo {author}
  {\bibfnamefont {S.}~\bibnamefont {Middey}}, \bibinfo {author} {\bibfnamefont
  {D.}~\bibnamefont {Meyers}}, \bibinfo {author} {\bibfnamefont
  {M.}~\bibnamefont {Kareev}}, \bibinfo {author} {\bibfnamefont
  {P.}~\bibnamefont {Shafer}}, \bibinfo {author} {\bibfnamefont
  {E.}~\bibnamefont {Arenholz}}, \ and\ \bibinfo {author} {\bibfnamefont
  {J.}~\bibnamefont {Chakhalian}},\ }\href@noop {} {\bibfield  {journal}
  {\bibinfo  {journal} {Applied Physics Letters}\ }\textbf {\bibinfo {volume}
  {109}},\ \bibinfo {pages} {152905} (\bibinfo {year}
  {2016}{\natexlab{a}})}\BibitemShut {NoStop}%
\bibitem [{\citenamefont {Liu}\ \emph {et~al.}(2015)\citenamefont {Liu},
  \citenamefont {Dean}, \citenamefont {Liu}, \citenamefont {Chiuzb{\u{a}}ian},
  \citenamefont {Jaouen}, \citenamefont {Nicolaou}, \citenamefont {Yin},
  \citenamefont {Serrao}, \citenamefont {Ramesh}, \citenamefont {Ding} \emph
  {et~al.}}]{liu2015probing}%
  \BibitemOpen
  \bibfield  {author} {\bibinfo {author} {\bibfnamefont {X.}~\bibnamefont
  {Liu}}, \bibinfo {author} {\bibfnamefont {M.}~\bibnamefont {Dean}}, \bibinfo
  {author} {\bibfnamefont {J.}~\bibnamefont {Liu}}, \bibinfo {author}
  {\bibfnamefont {S.}~\bibnamefont {Chiuzb{\u{a}}ian}}, \bibinfo {author}
  {\bibfnamefont {N.}~\bibnamefont {Jaouen}}, \bibinfo {author} {\bibfnamefont
  {A.}~\bibnamefont {Nicolaou}}, \bibinfo {author} {\bibfnamefont
  {W.}~\bibnamefont {Yin}}, \bibinfo {author} {\bibfnamefont {C.~R.}\
  \bibnamefont {Serrao}}, \bibinfo {author} {\bibfnamefont {R.}~\bibnamefont
  {Ramesh}}, \bibinfo {author} {\bibfnamefont {H.}~\bibnamefont {Ding}},  \emph
  {et~al.},\ }\href@noop {} {\bibfield  {journal} {\bibinfo  {journal} {Journal
  of Physics: Condensed Matter}\ }\textbf {\bibinfo {volume} {27}},\ \bibinfo
  {pages} {202202} (\bibinfo {year} {2015})}\BibitemShut {NoStop}%
\bibitem [{\citenamefont {Ament}\ \emph
  {et~al.}(2011{\natexlab{b}})\citenamefont {Ament}, \citenamefont {van
  Veenendaal}, \citenamefont {Devereaux}, \citenamefont {Hill},\ and\
  \citenamefont {van~den Brink}}]{ament2011resonant}%
  \BibitemOpen
  \bibfield  {author} {\bibinfo {author} {\bibfnamefont {L.~J.~P.}\
  \bibnamefont {Ament}}, \bibinfo {author} {\bibfnamefont {M.}~\bibnamefont
  {van Veenendaal}}, \bibinfo {author} {\bibfnamefont {T.~P.}\ \bibnamefont
  {Devereaux}}, \bibinfo {author} {\bibfnamefont {J.~P.}\ \bibnamefont {Hill}},
  \ and\ \bibinfo {author} {\bibfnamefont {J.}~\bibnamefont {van~den Brink}},\
  }\href@noop {} {\bibfield  {journal} {\bibinfo  {journal} {Rev. Mod. Phys.}\
  }\textbf {\bibinfo {volume} {83}},\ \bibinfo {pages} {705} (\bibinfo {year}
  {2011}{\natexlab{b}})}\BibitemShut {NoStop}%
\bibitem [{Note2()}]{Note2}%
  \BibitemOpen
  \bibinfo {note} {To ensure the change in relative intensity of the phonon
  features is not an artifact of this change, we present the normalized spectra
  with the fit elastic feature subtracted in the supplemental \cite
  {supplemental}.}\BibitemShut {Stop}%
\bibitem [{Note3()}]{Note3}%
  \BibitemOpen
  \bibinfo {note} {Here we use $j$ and $k$ instead of the usual $m$ and $n$ to
  avoid confusion with the superlattice layering indicies.}\BibitemShut {Stop}%
\bibitem [{\citenamefont {Iguchi}\ \emph {et~al.}(1992)\citenamefont {Iguchi},
  \citenamefont {Tamenori},\ and\ \citenamefont
  {Kubota}}]{Iguchi_TheoryBTO_EPC}%
  \BibitemOpen
  \bibfield  {author} {\bibinfo {author} {\bibfnamefont {E.}~\bibnamefont
  {Iguchi}}, \bibinfo {author} {\bibfnamefont {A.}~\bibnamefont {Tamenori}}, \
  and\ \bibinfo {author} {\bibfnamefont {N.}~\bibnamefont {Kubota}},\ }\href
  {\doibase 10.1103/PhysRevB.45.697} {\bibfield  {journal} {\bibinfo  {journal}
  {Phys. Rev. B}\ }\textbf {\bibinfo {volume} {45}},\ \bibinfo {pages} {697}
  (\bibinfo {year} {1992})}\BibitemShut {NoStop}%
\bibitem [{\citenamefont {Wang}\ \emph {et~al.}(2016)\citenamefont {Wang},
  \citenamefont {Walker}, \citenamefont {Tamai}, \citenamefont {Wang},
  \citenamefont {Ristic}, \citenamefont {Bruno}, \citenamefont {De~La~Torre},
  \citenamefont {Ricc{\`o}}, \citenamefont {Plumb}, \citenamefont {Shi} \emph
  {et~al.}}]{wang2016tailoring}%
  \BibitemOpen
  \bibfield  {author} {\bibinfo {author} {\bibfnamefont {Z.}~\bibnamefont
  {Wang}}, \bibinfo {author} {\bibfnamefont {S.~M.}\ \bibnamefont {Walker}},
  \bibinfo {author} {\bibfnamefont {A.}~\bibnamefont {Tamai}}, \bibinfo
  {author} {\bibfnamefont {Y.}~\bibnamefont {Wang}}, \bibinfo {author}
  {\bibfnamefont {Z.}~\bibnamefont {Ristic}}, \bibinfo {author} {\bibfnamefont
  {F.~Y.}\ \bibnamefont {Bruno}}, \bibinfo {author} {\bibfnamefont
  {A.}~\bibnamefont {De~La~Torre}}, \bibinfo {author} {\bibfnamefont
  {S.}~\bibnamefont {Ricc{\`o}}}, \bibinfo {author} {\bibfnamefont
  {N.}~\bibnamefont {Plumb}}, \bibinfo {author} {\bibfnamefont
  {M.}~\bibnamefont {Shi}},  \emph {et~al.},\ }\href@noop {} {\bibfield
  {journal} {\bibinfo  {journal} {Nature materials}\ }\textbf {\bibinfo
  {volume} {15}},\ \bibinfo {pages} {835} (\bibinfo {year} {2016})}\BibitemShut
  {NoStop}%
\bibitem [{\citenamefont {Gray}\ \emph {et~al.}(2016)\citenamefont {Gray},
  \citenamefont {Middey}, \citenamefont {Conti}, \citenamefont {Gray},
  \citenamefont {Kuo}, \citenamefont {Kaiser}, \citenamefont {Ueda},
  \citenamefont {Kobayashi}, \citenamefont {Meyers}, \citenamefont {Kareev}
  \emph {et~al.}}]{gray2016superconductor}%
  \BibitemOpen
  \bibfield  {author} {\bibinfo {author} {\bibfnamefont {B.}~\bibnamefont
  {Gray}}, \bibinfo {author} {\bibfnamefont {S.}~\bibnamefont {Middey}},
  \bibinfo {author} {\bibfnamefont {G.}~\bibnamefont {Conti}}, \bibinfo
  {author} {\bibfnamefont {A.}~\bibnamefont {Gray}}, \bibinfo {author}
  {\bibfnamefont {C.-T.}\ \bibnamefont {Kuo}}, \bibinfo {author} {\bibfnamefont
  {A.}~\bibnamefont {Kaiser}}, \bibinfo {author} {\bibfnamefont
  {S.}~\bibnamefont {Ueda}}, \bibinfo {author} {\bibfnamefont {K.}~\bibnamefont
  {Kobayashi}}, \bibinfo {author} {\bibfnamefont {D.}~\bibnamefont {Meyers}},
  \bibinfo {author} {\bibfnamefont {M.}~\bibnamefont {Kareev}},  \emph
  {et~al.},\ }\href@noop {} {\bibfield  {journal} {\bibinfo  {journal}
  {Scientific reports}\ }\textbf {\bibinfo {volume} {6}},\ \bibinfo {pages}
  {33184} (\bibinfo {year} {2016})}\BibitemShut {NoStop}%
\bibitem [{\citenamefont {Cao}\ \emph {et~al.}(2016{\natexlab{b}})\citenamefont
  {Cao}, \citenamefont {Liu}, \citenamefont {Kareev}, \citenamefont
  {Choudhury}, \citenamefont {Middey}, \citenamefont {Meyers}, \citenamefont
  {Kim}, \citenamefont {Ryan}, \citenamefont {Freeland},\ and\ \citenamefont
  {Chakhalian}}]{cao2016engineered}%
  \BibitemOpen
  \bibfield  {author} {\bibinfo {author} {\bibfnamefont {Y.}~\bibnamefont
  {Cao}}, \bibinfo {author} {\bibfnamefont {X.}~\bibnamefont {Liu}}, \bibinfo
  {author} {\bibfnamefont {M.}~\bibnamefont {Kareev}}, \bibinfo {author}
  {\bibfnamefont {D.}~\bibnamefont {Choudhury}}, \bibinfo {author}
  {\bibfnamefont {S.}~\bibnamefont {Middey}}, \bibinfo {author} {\bibfnamefont
  {D.}~\bibnamefont {Meyers}}, \bibinfo {author} {\bibfnamefont {J.-W.}\
  \bibnamefont {Kim}}, \bibinfo {author} {\bibfnamefont {P.}~\bibnamefont
  {Ryan}}, \bibinfo {author} {\bibfnamefont {J.}~\bibnamefont {Freeland}}, \
  and\ \bibinfo {author} {\bibfnamefont {J.}~\bibnamefont {Chakhalian}},\
  }\href@noop {} {\bibfield  {journal} {\bibinfo  {journal} {Nature
  communications}\ }\textbf {\bibinfo {volume} {7}} (\bibinfo {year}
  {2016}{\natexlab{b}})}\BibitemShut {NoStop}%
\bibitem [{\citenamefont {Zhong}\ and\ \citenamefont
  {Hansmann}(2017)}]{zhong2017band}%
  \BibitemOpen
  \bibfield  {author} {\bibinfo {author} {\bibfnamefont {Z.}~\bibnamefont
  {Zhong}}\ and\ \bibinfo {author} {\bibfnamefont {P.}~\bibnamefont
  {Hansmann}},\ }\href@noop {} {\bibfield  {journal} {\bibinfo  {journal}
  {Physical Review X}\ }\textbf {\bibinfo {volume} {7}},\ \bibinfo {pages}
  {011023} (\bibinfo {year} {2017})}\BibitemShut {NoStop}%
\bibitem [{\citenamefont {Hwang}\ \emph {et~al.}(2012)\citenamefont {Hwang},
  \citenamefont {Iwasa}, \citenamefont {Kawasaki}, \citenamefont {Keimer},
  \citenamefont {Nagaosa},\ and\ \citenamefont {Tokura}}]{Hwang_ReviewSL2012}%
  \BibitemOpen
  \bibfield  {author} {\bibinfo {author} {\bibfnamefont {H.~Y.}\ \bibnamefont
  {Hwang}}, \bibinfo {author} {\bibfnamefont {Y.}~\bibnamefont {Iwasa}},
  \bibinfo {author} {\bibfnamefont {M.}~\bibnamefont {Kawasaki}}, \bibinfo
  {author} {\bibfnamefont {B.}~\bibnamefont {Keimer}}, \bibinfo {author}
  {\bibfnamefont {N.}~\bibnamefont {Nagaosa}}, \ and\ \bibinfo {author}
  {\bibfnamefont {Y.}~\bibnamefont {Tokura}},\ }\href {\doibase
  10.1038/nmat3223} {\bibfield  {journal} {\bibinfo  {journal} {Nat. Mater.}\
  }\textbf {\bibinfo {volume} {11}},\ \bibinfo {pages} {103} (\bibinfo {year}
  {2012})}\BibitemShut {NoStop}%
\bibitem [{\citenamefont {Chakhalian}\ \emph {et~al.}(2014)\citenamefont
  {Chakhalian}, \citenamefont {Freeland}, \citenamefont {Millis}, \citenamefont
  {Panagopoulos},\ and\ \citenamefont {Rondinelli}}]{Chakhalian_ReviewSL2014}%
  \BibitemOpen
  \bibfield  {author} {\bibinfo {author} {\bibfnamefont {J.}~\bibnamefont
  {Chakhalian}}, \bibinfo {author} {\bibfnamefont {J.~W.}\ \bibnamefont
  {Freeland}}, \bibinfo {author} {\bibfnamefont {A.~J.}\ \bibnamefont
  {Millis}}, \bibinfo {author} {\bibfnamefont {C.}~\bibnamefont
  {Panagopoulos}}, \ and\ \bibinfo {author} {\bibfnamefont {J.~M.}\
  \bibnamefont {Rondinelli}},\ }\href {\doibase 10.1103/RevModPhys.86.1189}
  {\bibfield  {journal} {\bibinfo  {journal} {Rev. Mod. Phys.}\ }\textbf
  {\bibinfo {volume} {86}},\ \bibinfo {pages} {1189} (\bibinfo {year}
  {2014})}\BibitemShut {NoStop}%
\bibitem [{\citenamefont {Mannhart}\ and\ \citenamefont
  {Schlom}(2010)}]{mannhart2010oxide}%
  \BibitemOpen
  \bibfield  {author} {\bibinfo {author} {\bibfnamefont {J.}~\bibnamefont
  {Mannhart}}\ and\ \bibinfo {author} {\bibfnamefont {D.}~\bibnamefont
  {Schlom}},\ }\href@noop {} {\bibfield  {journal} {\bibinfo  {journal}
  {Science}\ }\textbf {\bibinfo {volume} {327}},\ \bibinfo {pages} {1607}
  (\bibinfo {year} {2010})}\BibitemShut {NoStop}%
\bibitem [{\citenamefont {Matsuno}\ \emph {et~al.}(2015)\citenamefont
  {Matsuno}, \citenamefont {Ihara}, \citenamefont {Yamamura}, \citenamefont
  {Wadati}, \citenamefont {Ishii}, \citenamefont {Shankar}, \citenamefont
  {Kee},\ and\ \citenamefont {Takagi}}]{Matsuno2015_SIOSTO}%
  \BibitemOpen
  \bibfield  {author} {\bibinfo {author} {\bibfnamefont {J.}~\bibnamefont
  {Matsuno}}, \bibinfo {author} {\bibfnamefont {K.}~\bibnamefont {Ihara}},
  \bibinfo {author} {\bibfnamefont {S.}~\bibnamefont {Yamamura}}, \bibinfo
  {author} {\bibfnamefont {H.}~\bibnamefont {Wadati}}, \bibinfo {author}
  {\bibfnamefont {K.}~\bibnamefont {Ishii}}, \bibinfo {author} {\bibfnamefont
  {V.~V.}\ \bibnamefont {Shankar}}, \bibinfo {author} {\bibfnamefont {H.-Y.}\
  \bibnamefont {Kee}}, \ and\ \bibinfo {author} {\bibfnamefont
  {H.}~\bibnamefont {Takagi}},\ }\href {\doibase
  10.1103/PhysRevLett.114.247209} {\bibfield  {journal} {\bibinfo  {journal}
  {Phys. Rev. Lett.}\ }\textbf {\bibinfo {volume} {114}},\ \bibinfo {pages}
  {247209} (\bibinfo {year} {2015})}\BibitemShut {NoStop}%
\bibitem [{\citenamefont {Fan}\ and\ \citenamefont
  {Yunoki}(2015)}]{fan2015_LDA}%
  \BibitemOpen
  \bibfield  {author} {\bibinfo {author} {\bibfnamefont {W.}~\bibnamefont
  {Fan}}\ and\ \bibinfo {author} {\bibfnamefont {S.}~\bibnamefont {Yunoki}},\
  }\href {\doibase https://doi.org/10.1088/1742-6596/592/1/012139} {\bibfield
  {journal} {\bibinfo  {journal} {J. Phys. Conf. Ser.}\ }\textbf {\bibinfo
  {volume} {592}},\ \bibinfo {pages} {012139} (\bibinfo {year}
  {2015})}\BibitemShut {NoStop}%
\bibitem [{\citenamefont {Kim}\ \emph {et~al.}(2017)\citenamefont {Kim},
  \citenamefont {Liu},\ and\ \citenamefont {Franchini}}]{Kim2017_STSIOtheory}%
  \BibitemOpen
  \bibfield  {author} {\bibinfo {author} {\bibfnamefont {B.}~\bibnamefont
  {Kim}}, \bibinfo {author} {\bibfnamefont {P.}~\bibnamefont {Liu}}, \ and\
  \bibinfo {author} {\bibfnamefont {C.}~\bibnamefont {Franchini}},\ }\href
  {\doibase 10.1103/PhysRevB.95.115111} {\bibfield  {journal} {\bibinfo
  {journal} {Phys. Rev. B}\ }\textbf {\bibinfo {volume} {95}},\ \bibinfo
  {pages} {115111} (\bibinfo {year} {2017})}\BibitemShut {NoStop}%
\bibitem [{\citenamefont {Spinelli}\ \emph {et~al.}(2010)\citenamefont
  {Spinelli}, \citenamefont {Torija}, \citenamefont {Liu}, \citenamefont
  {Jan},\ and\ \citenamefont {Leighton}}]{spinelli2010electronic}%
  \BibitemOpen
  \bibfield  {author} {\bibinfo {author} {\bibfnamefont {A.}~\bibnamefont
  {Spinelli}}, \bibinfo {author} {\bibfnamefont {M.}~\bibnamefont {Torija}},
  \bibinfo {author} {\bibfnamefont {C.}~\bibnamefont {Liu}}, \bibinfo {author}
  {\bibfnamefont {C.}~\bibnamefont {Jan}}, \ and\ \bibinfo {author}
  {\bibfnamefont {C.}~\bibnamefont {Leighton}},\ }\href@noop {} {\bibfield
  {journal} {\bibinfo  {journal} {Physical Review B}\ }\textbf {\bibinfo
  {volume} {81}},\ \bibinfo {pages} {155110} (\bibinfo {year}
  {2010})}\BibitemShut {NoStop}%
\bibitem [{\citenamefont {Moos}\ and\ \citenamefont
  {H{\"a}rdtl}(1996)}]{moos1996electronic}%
  \BibitemOpen
  \bibfield  {author} {\bibinfo {author} {\bibfnamefont {R.}~\bibnamefont
  {Moos}}\ and\ \bibinfo {author} {\bibfnamefont {K.~H.}\ \bibnamefont
  {H{\"a}rdtl}},\ }\href@noop {} {\bibfield  {journal} {\bibinfo  {journal}
  {Journal of Applied Physics}\ }\textbf {\bibinfo {volume} {80}},\ \bibinfo
  {pages} {393} (\bibinfo {year} {1996})}\BibitemShut {NoStop}%
\bibitem [{\citenamefont {Kim}\ \emph {et~al.}(2014)\citenamefont {Kim},
  \citenamefont {Kim},\ and\ \citenamefont {Han}}]{khkim_2014}%
  \BibitemOpen
  \bibfield  {author} {\bibinfo {author} {\bibfnamefont {K.-H.}\ \bibnamefont
  {Kim}}, \bibinfo {author} {\bibfnamefont {H.-S.}\ \bibnamefont {Kim}}, \ and\
  \bibinfo {author} {\bibfnamefont {M.~J.}\ \bibnamefont {Han}},\ }\href
  {http://stacks.iop.org/0953-8984/26/i=18/a=185501} {\bibfield  {journal}
  {\bibinfo  {journal} {Journal of Physics: Condensed Matter}\ }\textbf
  {\bibinfo {volume} {26}},\ \bibinfo {pages} {185501} (\bibinfo {year}
  {2014})}\BibitemShut {NoStop}%
\bibitem [{\citenamefont {Fr\"{o}hlich}(1954)}]{frohlich_1954}%
  \BibitemOpen
  \bibfield  {author} {\bibinfo {author} {\bibfnamefont {H.}~\bibnamefont
  {Fr\"{o}hlich}},\ }\href {\doibase 10.1080/00018735400101213} {\bibfield
  {journal} {\bibinfo  {journal} {Advances in Physics}\ }\textbf {\bibinfo
  {volume} {3}} (\bibinfo {year} {1954}),\ 10.1080/00018735400101213},\ \Eprint
  {http://arxiv.org/abs/https://doi.org/10.1080/00018735400101213}
  {https://doi.org/10.1080/00018735400101213} \BibitemShut {NoStop}%
\bibitem [{\citenamefont {Matthias}(1949)}]{matthias_1949}%
  \BibitemOpen
  \bibfield  {author} {\bibinfo {author} {\bibfnamefont {B.~T.}\ \bibnamefont
  {Matthias}},\ }\href {\doibase 10.1103/PhysRev.75.1771} {\bibfield  {journal}
  {\bibinfo  {journal} {Phys. Rev.}\ }\textbf {\bibinfo {volume} {75}},\
  \bibinfo {pages} {1771} (\bibinfo {year} {1949})}\BibitemShut {NoStop}%
\bibitem [{\citenamefont {Hill}(2000)}]{nicola_2000}%
  \BibitemOpen
  \bibfield  {author} {\bibinfo {author} {\bibfnamefont {N.~A.}\ \bibnamefont
  {Hill}},\ }\href {\doibase 10.1021/jp000114x} {\bibfield  {journal} {\bibinfo
   {journal} {The Journal of Physical Chemistry B}\ }\textbf {\bibinfo {volume}
  {104}},\ \bibinfo {pages} {6694} (\bibinfo {year} {2000})},\ \Eprint
  {http://arxiv.org/abs/https://doi.org/10.1021/jp000114x}
  {https://doi.org/10.1021/jp000114x} \BibitemShut {NoStop}%
\bibitem [{\citenamefont {Giustino}(2017{\natexlab{b}})}]{giustino_rmp_2017}%
  \BibitemOpen
  \bibfield  {author} {\bibinfo {author} {\bibfnamefont {F.}~\bibnamefont
  {Giustino}},\ }\href {\doibase 10.1103/RevModPhys.89.015003} {\bibfield
  {journal} {\bibinfo  {journal} {Rev. Mod. Phys.}\ }\textbf {\bibinfo {volume}
  {89}},\ \bibinfo {pages} {015003} (\bibinfo {year}
  {2017}{\natexlab{b}})}\BibitemShut {NoStop}%
\bibitem [{\citenamefont {Klimin}\ \emph {et~al.}(2017)\citenamefont {Klimin},
  \citenamefont {Tempere}, \citenamefont {Devreese},\ and\ \citenamefont {der
  Marel}}]{Klimin2017}%
  \BibitemOpen
  \bibfield  {author} {\bibinfo {author} {\bibfnamefont {S.~N.}\ \bibnamefont
  {Klimin}}, \bibinfo {author} {\bibfnamefont {J.}~\bibnamefont {Tempere}},
  \bibinfo {author} {\bibfnamefont {J.~T.}\ \bibnamefont {Devreese}}, \ and\
  \bibinfo {author} {\bibfnamefont {D.~v.}\ \bibnamefont {der Marel}},\ }\href
  {\doibase 10.1007/s10948-016-3664-2} {\bibfield  {journal} {\bibinfo
  {journal} {Journal of Superconductivity and Novel Magnetism}\ }\textbf
  {\bibinfo {volume} {30}},\ \bibinfo {pages} {757} (\bibinfo {year}
  {2017})}\BibitemShut {NoStop}%
\end{thebibliography}
\end{document}